\documentclass[12pt]{article}

\usepackage{scicite}

\usepackage{times}

\usepackage{graphicx}
\usepackage{journals_science}

\newenvironment{sciabstract}{%
\begin{quote} \bf}
{\end{quote}}

\renewcommand\refname{References and Notes}

\newcounter{lastnote}
\newenvironment{scilastnote}{%
\setcounter{lastnote}{\value{enumiv}}%
\addtocounter{lastnote}{+1}%
\begin{list}%
{\arabic{lastnote}.}
{\setlength{\leftmargin}{.22in}}
{\setlength{\labelsep}{.5em}}}
{\end{list}}

\title{A Population of Compact Elliptical Galaxies Detected with the Virtual
Observatory}

\author
{Igor Chilingarian$^{1,2,3\ast}$, V\'eronique Cayatte$^{4}$, Yves Revaz$^{5}$, \\
Serguei Dodonov$^{6}$, Daniel Durand$^{7}$, Florence Durret $^{8,9}$, \\
Alberto Micol$^{10}$ \& Eric Slezak$^{11}$\\
\\
\normalsize{$^{1}$ Observatoire de Paris-Meudon, LERMA, UMR~8112, }\\
\normalsize{61 Av. de l'Observatoire, 75014 Paris, France}\\
\normalsize{$^{2}$ Sternberg Astronomical Institute, Moscow State University,  }\\
\normalsize{13 Universitetsky prospect, Moscow, 119296, Russia}\\
\normalsize{$^{3}$ Observatoire de Paris, VO Paris Data Centre,  }\\
\normalsize{61 Av. de l'Observatoire, 75014 Paris, France}\\
\normalsize{$^{4}$ Observatoire de Paris-Meudon, LUTH, UMR~8102,  }\\
\normalsize{5 pl. Jules Janssen, 92195 Meudon, France}\\
\normalsize{$^{5}$ Ecole Polytechnique F\'ed\'erale de Lausanne, Laboratory of Astrophysics,  }\\
\normalsize{51. Ch. des Maillettes, 1290 Sauverny, Switzerland}\\
\normalsize{$^{6}$ Special Astrophysical Observatory, Nizhnij Arkhyz,  }\\
\normalsize{Zelenchukskij region, Karachai-Cirkassian Republic, 369167, Russia}\\
\normalsize{$^{7}$ National Research Council Canada, Herzberg Institute of Astrophysics,  }\\
\normalsize{5071 W. Saanich Rd, Victoria, BC, Canada V9E 2E7}\\
\normalsize{$^{8}$ CNRS, UMR~7095, Institut d'astrophysique de Paris,  }\\
\normalsize{$^{9}$ UPMC Universit\'e Paris 06, UMR~7095, Institut d'Astrophysique de Paris, }\\
\normalsize{98bis, boulevard Arago, 75014 Paris, France}\\
\normalsize{$^{10}$ European Southern Observatory,  }\\
\normalsize{Karl-Schwarzschild-Strasse 2, 85748 Garching bei Munchen, Germany}\\
\normalsize{$^{11}$ University of Nice Sophia Antipolis, CNRS, Observatoire de la C\^ote d'Azur,  }\\
\normalsize{B.P. 4229, 06304 Nice Cedex 4, France}\\
\\
\normalsize{$^\ast$To whom correspondence should be addressed; E-mail:
igor.chilingarian@obspm.fr}
}

\date{}

\begin{document} 


\baselineskip24pt


\maketitle


\begin{sciabstract}
Compact elliptical galaxies are characterized by small sizes and high
stellar densities. They are thought to form through tidal stripping of
massive progenitors. However, only a handful of them were known,
preventing us from understanding the role played by this mechanism in galaxy
evolution. We present a population of 21 compact elliptical galaxies
gathered with the Virtual Observatory. Follow-up spectroscopy and data
mining, using high resolution images and large databases, show that all the
galaxies exhibit old metal-rich stellar populations different from those of
dwarf elliptical galaxies of similar masses but similar to those of more
massive early-type galaxies, supporting the tidal stripping scenario. Their
internal properties are reproduced by numerical simulations, which result in
compact dynamically hot remnants resembling the galaxies in our sample.
\end{sciabstract}

Present-day clusters of galaxies host rich populations of dwarf elliptical
(dE) and lenticular (dS0) galaxies\cite{FB94} having regular morphology and
lacking ongoing star formation and inter-stellar medium (ISM). These
galaxies are thought to form by internal processes such as supernova
feedback to the star formation\cite{DS86}, or external agents [ram pressure
stripping by hot intergalactic gas\cite{GG72} and/or gravitational
harassment\cite{Moore+96}] acting on gas-rich progenitors. Tidal stripping
had not been considered as an important mechanism governing galaxy formation
until the recent discovery of ultra-compact dwarf
galaxies\cite{MHI02,Drinkwater+03} (UCD), i.e. very compact stellar systems
several times more massive than known globular clusters. However, UCDs
($L \sim 10^7 L_{\odot}$) are about two orders of magnitude less luminous
than bright dEs and, therefore, can be studied in
only a handful of nearby galaxy clusters.

Compact elliptical (cE), or $M32$-like galaxies, which are also thought to
form through tidal stripping \cite{BCDG01}, have luminosities ($\sim
10^9 L_{\odot}$ \cite{Graham02,Mieske+05,Chilingarian+07,Price+09}) similar to
those of dEs, whereas their half-light radii ($R_{e} \sim 0.25$~kpc) are
several times smaller resulting in much higher mean surface brightness
($\langle \mu \rangle_{e}$) values compared to dEs. These two criteria are
easy to formalize and apply to members of nearby galaxy clusters at known
distances, hence having a known spatial scale. Ground-based
optical telescopes cannot resolve objects the size of 0.25~kpc beyond
$50$~Mpc. To find them in clusters out to 200~Mpc, thus increasing by a
factor of 60 the volume of the nearby Universe where cEs remain spatially
resolved, it is necessary to use the Hubble Space Telescope (HST).

We created a workflow, i.e. an automatic data retrieval and analysis system
to search for cE galaxies in large data collections provided by the Virtual
Observatory (VO, \cite{VOdef}). It comprised the following steps. (i)
Identify nearby galaxy clusters at redshifts $z < 0.055$ using the Vizier
Service \cite{Vizier} at the Centre de Donn\'ees Astronomiques de
Strasbourg. Without this condition our potential candidate cE galaxies would
have been too faint for spectroscopic follow-up. (ii) Once the sources were
identified, gather more precise measurements using other VO services
including the NASA/IPAC Extragalactic Database (NED, \cite{NED}). (iii) Use
the IVOA Simple Image Access Protocol to find and fetch the HST images of
selected galaxy clusters from the Hubble Legacy Archive (HLA, \cite{HLA}).
(iv) For each image, run the SExtractor source identification
software\cite{BA96} to obtain half-light radii, total luminosities and
approximate light profiles for all galaxies in each frame. (v) Apply color
corrections to homogenize the results for all photometric bands and use the
surface brightness -- half-light radius criteria to find cE candidates. (vi)
Finally, query NED, Vizier and a database of Sloan Digital Sky Survey Data
Release 7\cite{SDSS_DR7} to find additional information for candidate
objects, such as published redshifts and integrated photometry.

Having applied the workflow to the entire HST Wide-Field Planetary Camera-2
(WFPC2) data collection in the HLA we ended up with archival images of 63
clusters with 55 candidate cE and tidally stripped elliptical galaxies in 26
of them. 


Our workflow, which uses only imaging data, may confuse cE galaxies with (i)
foreground or cluster compact star-forming galaxies; (ii) background giant
early-type galaxies hosting bright active nuclei (AGN); (iii) background
post-starburst galaxies.  Star-forming galaxies can be discriminated
automatically through their blue colors if multi-band data are available, or
by eye, examining their clumpy morphology. The two other cases may arise
when the distance to the background object is 2--3 times the distance to the
cluster, i.e. $<$400--500~Mpc ($z<0.12$) in our study. The probability of
having a post-starburst galaxy at this redshift is very low \cite{Goto+03}
and AGNs can be ruled out by checking X-ray point source catalogues.

We immediately confirmed the nature of 14 cE galaxies (see Fig.~1). For 6 of
them, where archival SDSS DR7 spectra were available, we derived their
physical properties: internal kinematics and stellar populations. For the
other 8 candidate galaxies, we obtained redshifts from the literature through
the NED and Vizier services.

To increase the sample of confirmed cE galaxies, we observed three of the
galaxy clusters, Abell~160, Abell~189, and Abell~397, hosting 7 candidate
galaxies altogether having $2.3\cdot10^{8}<L_{B}<1.1\cdot10^{9} L_{\odot}$.
This was done with the multi-slit unit of the SCORPIO spectrograph \cite{AM05} at the
Russian 6-m ``Big Telescope Alt-azimuthal'' in August 2008 (SOM). The
spectra were fit with high-resolution stellar population
models using the {\sc NBursts} full spectral fitting technique \cite{CPSK07}
(SOM), obtaining precise radial velocities, internal velocity dispersions,
luminosity-weighted stellar ages and metallicities. All 7 candidate objects
were confirmed to be cluster members having global internal velocity
dispersions between 50 and 100~km~s$^{-1}$ and old stellar populations. 
One of them exhibits a possible tidal feature (Fig.~2).


None of the 21 cE galaxies in our sample (summarized in Tables~S1,~S2)
exhibits a young stellar population. The luminosity-weighted metallicities
are between $Z_{\odot}/2.5$ ($[$Fe/H$] = -0.4$~dex) and the solar value for
all galaxies except for ACO~189~J012325.96+014236.2 where it is as low as
$Z_{\odot}/5$. Three galaxies are offset by $> +0.3$~dex from the
luminosity-metallicity relation for early-type galaxies (Fig.~1, bottom);
three others reside +0.2 to +0.3~dex above it. The internal velocity
dispersions are close to the values expected for the Faber-Jackson
\cite{FJ76} relation and systematically above those of dEs. All these
properties support the scenario where cE galaxies are created through tidal
stripping of intermediate-luminosity disc galaxies.

Tidal stripping of an early-type disc galaxy should primarily affect its
extended disc. Its centrally concentrated bulge should be able to survive
even severe tidal interactions. After dynamical relaxation, a
tidally-stripped remnant may slightly expand, decreasing its stellar
velocity dispersion. For ISM-deficient galaxies, such as early-type systems
in clusters, tidal stripping will not change the overall age and metallicity
of their stars, but the total stellar mass will be significantly reduced.
Because there is a relation between mean metallicities and stellar masses of
galaxies \cite{Gallazzi+05,Renzini06}, this process will create objects with
uncommon stellar populations; i.e. systems that are too metal-rich 
for their observed stellar masses, corresponding to what we see in galaxies
populating the central region of the Abell~496 cluster\cite{Chilingarian+08}
and in UCDs\cite{CCB08}.

The stellar metallicity can also be increased by star formation induced by
external sources, e.g. by mergers or encounters with gas-rich galaxies
where tidal compression may significantly increase the star formation
efficiency \cite{EvdV08}. However, in this case the mean age of the stellar
population would decrease because newly formed stars have much lower
mass-to-light ratios in the optical than older populations do, contributing
significantly to the total light, even with low mass fractions. All the
galaxies in our sample exhibit old populations, hence ruling out this
scenario.

Tidal effects must be stronger in the vicinities of massive galaxies, in
particular around cluster dominant galaxies (cD). 
%

To study the efficiency of tidal stripping we simulated the interaction of a
disc galaxy with a galaxy cluster potential using the {\sc gadget-2}
code\cite{Springel05} (SOM for further details). We explored the tidal
stripping effects at a spatial resolution ($\sim 100$~pc) several times
higher than that of typical simulations of galaxy clusters, checking 32
different orbital configurations. We converted the stellar density maps
produced by our simulations into surface brightness and processed them using
the SExtractor software in the same way as we did for the search of cE galaxies
in HST images.

Our simulations demonstrate the efficiency of tidal stripping in reducing
the stellar mass of a disc galaxy. Even in case of quasi-circular orbital
configurations, the large-scale stellar disc was heavily stripped,
decreasing the galaxy stellar mass by a factor of 2 (Fig.~3) on a
timescale of 600--700~Myr. We compared the evolution of total magnitude,
surface brightness and internal velocity dispersion of a stripped galaxy to
observations (Fig.~1, top and middle panels). Interactions on radial orbits
resulted in heavier mass loss of up-to 90 per cent, although a remnant
became quickly accreted by a cD galaxy. Presumably, by scaling down the
systems in mass, i.e. by replacing a giant disc with a low-luminosity or
dwarf S0, it should be possible to reproduce the formation of UCD and
transitional cE/UCD\cite{CM08,Price+09} galaxies.

In our study we used Virtual Observatory data mining to convert
the class of cE galaxies from ``unique'' into ``common in certain
environmental conditions'', i.e. more frequent than was previously
thought. We confirmed the nature of 21 galaxies selected by the VO workflow
with follow-up observations and archival data. We also reproduced their properties
with numerical simulations. We can confirm that tidal stripping of the stellar
component plays an important role in the morphological transformation of
galaxies in dense environments, producing remnants spanning a luminosity
range of four orders of magnitude.

%


\begin{scilastnote}
\item This study is based on observations made with the
NASA/ESA Hubble Space Telescope, and obtained from the Hubble Legacy
Archive, which is a collaboration between the Space Telescope Science
Institute (STScI/NASA), the Space Telescope European Coordinating Facility
(ST-ECF/ESA) and the Canadian Astronomy Data Centre (CADC/NRC/CSA);
observations collected with the 6-m telescope of the Special Astrophysical
Observatory (SAO) of the Russian Academy of Sciences (RAS) operated under
the financial support of the Science Department of Russia (registration
number 01-43). The simulations were run on the Callisto cluster at
the \'Ecole Polytechnique F\'ed\'erale de Lausanne (EPFL) and on
the Regor cluster of the Geneva Observatory.
This research made use of SAOImage DS9 developed by
Smithsonian Astrophysical Observatory, Aladin developed by the Centre de
Donn\'ees Astronomiques de Strasbourg, ``exploresdss'' script by G.~Mamon;
the VizieR catalogue access tool, CDS, Strasbourg, France, and the NASA/IPAC
Extragalactic Database (NED) operated by the Jet Propulsion
Laboratory, California Institute of Technology, under contract with the
National Aeronautics and Space Administration. Funding for the SDSS and
SDSS-II has been provided by the Alfred P. Sloan Foundation, the
Participating Institutions, the National Science Foundation, the U.S.
Department of Energy, the National Aeronautics and Space Administration, the
Japanese Monbukagakusho, the Max Planck Society, and the Higher Education
Funding Council for England. The SDSS Web Site is http://www.sdss.org/. The
simulation data analysis and galaxy maps were done using the
parallelized Python pNbody package (http://obswww.unige.ch/$\sim$revaz/pNbody/). 
This work was supported by the Swiss National Science Foundation. IC
acknowledges additional support from the RFBR grant 07-02-00229-a. Special
thanks to F.~Combes and G.~Mamon for useful discussions and
suggestions and to R.~Trilling who kindly agreed to edit the manuscript.
The content of the workflow and its explicit description are available on
the web-pages of the VO Paris Data Centre (http://vo-web.obspm.fr/) and the
European Virtual Observatory EURO-VO.
\end{scilastnote}


\begin{figure}
\includegraphics{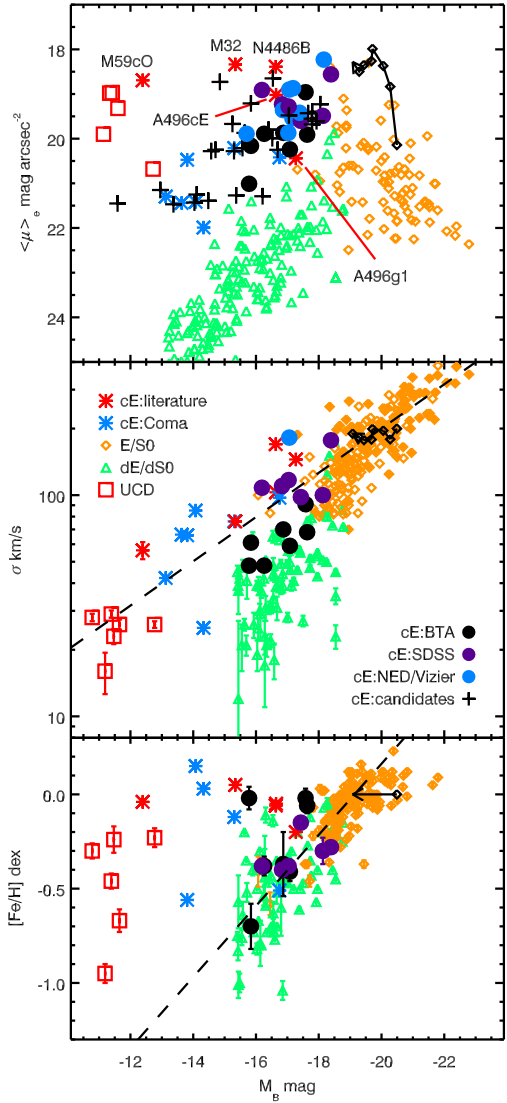}
\caption{Structure, dynamics and stellar populations of early-type
galaxies. Mean surface brightness within a half-light radius $\langle \mu
\rangle_{e}$ (top panel), global stellar velocity dispersion $\sigma$
(middle panel) and metallicity [Fe/H] (bottom panel) of galaxies are
displayed as a function of the $B$-band absolute magnitudes. The confirmed
tidally stripped galaxies are displayed with filled
circles of black, violet and blue colours for BTA-spectroscopy,
SDSS-spectroscopy, and literature redshifts from NED or Vizier,
respectively. Crosses denote the candidate cE galaxies without
spectroscopic confirmation. Red asterisks are for known compact elliptical
galaxies, blue asterisks are for recently discovered compact stellar systems
in the Coma cluster \cite{Price+09}. The dashed lines in the middle and 
bottom panel show the Faber-Jackson\cite{FJ76} and 
luminosity-metallicity\cite{CCB08} relations. Literature data (dE/dS0
galaxies\cite{BJ98,Chilingarian+08,Chilingarian09}, intermediate-luminosity
and giant early-type galaxies\cite{BBF92}, UCDs\cite{CCB08}) are
complemented with the data for 140 early-type galaxies in the Coma cluster
from the SDSS DR7 spectra obtained with the {\sc NBursts} technique (SOM).
All metallicity measurements shown in the bottom panel, except those of Coma 
compact stellar systems, are homogeneous as they were obtained using the 
same data analysis technique.
Black path displays the evolution of a simulated barred early-type spiral 
tidally stripped by the cluster potential on a quasi-circular orbit (SOM)
followed for 2~Gyr.}
\end{figure}

\begin{figure}
\includegraphics[width=12cm]{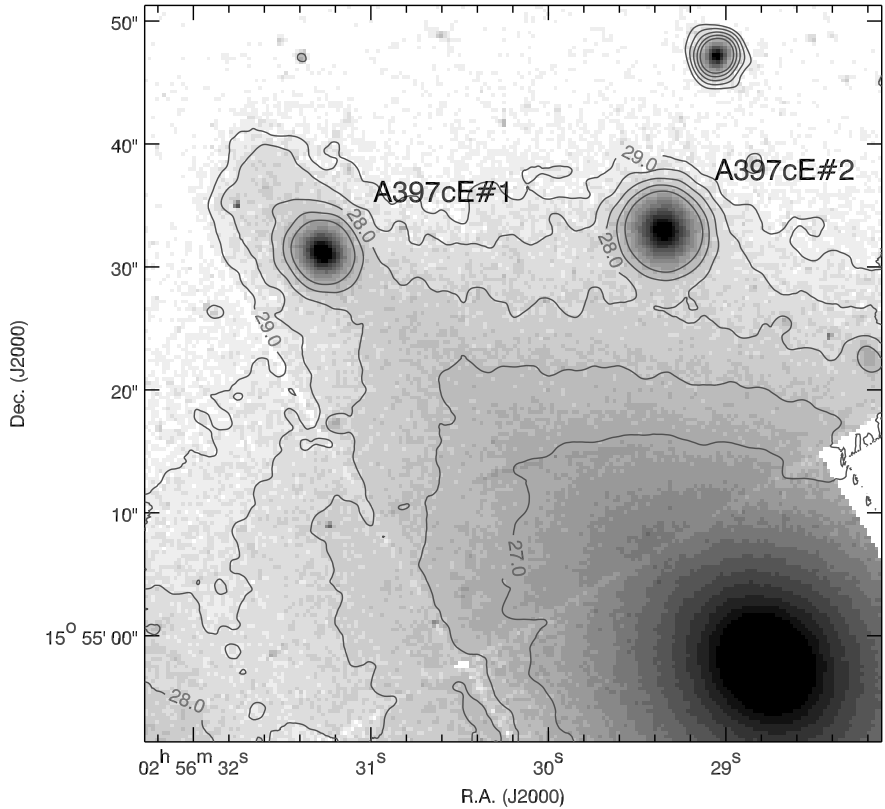}
\caption{Fragment of a WFPC2 HST image of
the central region of Abell~397 obtained from the HLA with two candidate cE
galaxies; north is up, east is left. The overplotted isophotes correspond to
surface brightness levels from 29.0 to 27.0~mag~arcsec$^{-2}$ in the $F814W$
band with 0.5~mag~arcsec$^{-2}$ interval. A397cE\#1 exhibits a
prominent extended low surface brightness feature toward the north-east,
probably originating from its tidal interaction with the cluster dominant
galaxy UGC~2413 in the lower-right corner.}
\end{figure}

\begin{figure}
\includegraphics{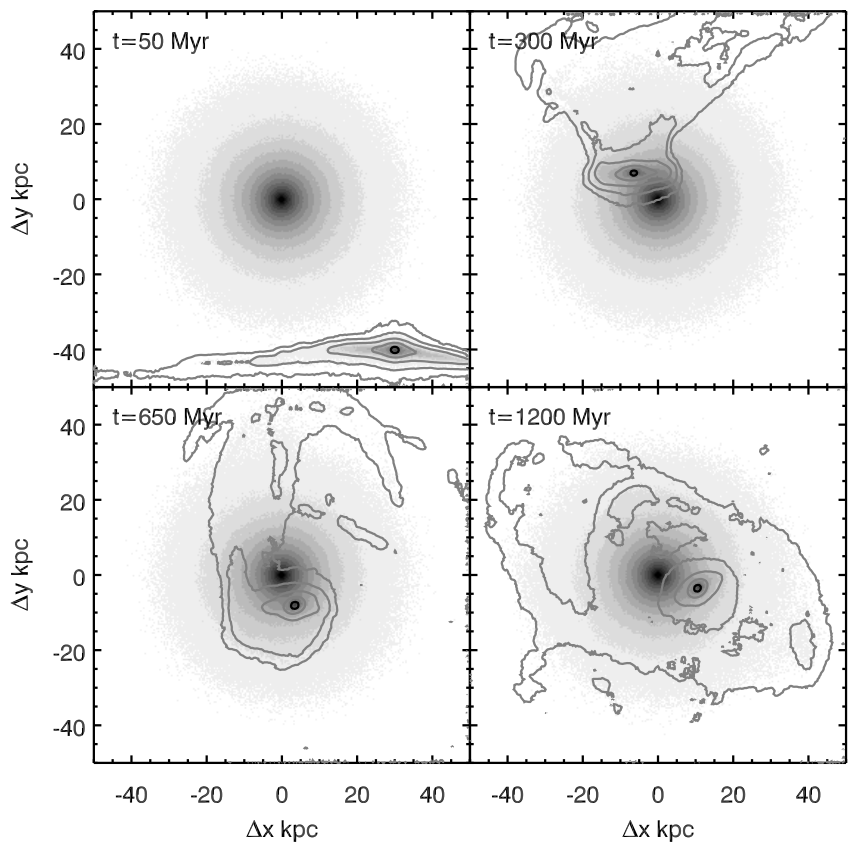}

\caption{Numerical simulations of the tidal stripping of an
intermediate-luminosity barred disc galaxy. The panels display four
snapshots of an $N$-body simulation of the tidal disruption of a barred disc
galaxy, initially on a quasi-circular orbit, by the cluster cD galaxy shown in
the centre. The disc galaxy plane is perpendicular to the initial orbital
plane. Each panel is 100$\times$100~kpc in size, the density scale is
logarithmic. The density contours (every 2~mag~arcsec$^{-2}$) are displayed
for the stars which initially belonged to the disc galaxy. The simulations
demonstrate how the outer disc of a galaxy gets stripped leaving a compact
remnant and stripped material in arcs and tails.}

\end{figure}

\clearpage
\renewcommand{\thefigure}{S\arabic{figure}}
\setcounter{figure}{0}

\renewcommand{\thetable}{S\arabic{table}}
\setcounter{table}{0}

\begin{center}
\begin{large}
{\bf Supporting online material}
\end{large}
\end{center}

\section*{Observations, data reduction and analysis}

We observed three galaxy clusters (Abell~160, Abell~189, and Abell~397),
hosting 7 candidate galaxies altogether ($-17.1<M_{B}<-15.4$~mag) with the
SCORPIO (Spectral Camera with Optical Reducer for Photometric and
Interferometric Observations) spectrograph \cite{AM05} at the Russian 6-m
``Big Telescope Alt-azimuthal'' during one observing run on the nights of
25th, 27th, and 28th of August 2008 (observational programme ``Kinematics
and stellar populations of newly discovered compact elliptical galaxies'',
P.I.:IC). We used the multi-slit mode of the instrument and the VPHG2300g
grism centered at 5250~\AA. This allowed us to observe up to 16 galaxies
simultaneously in every 6$\times$6~arcmin field of view, with a spectral
resolution of about $R=2200$ (FWHM = 2.4~\AA), using 1.2~arcsec-wide
18~arcsec-long interactively placed slitlets, controlled by electromagnets
in the focal plane of the focal reducer. The total exposure times were 3600,
6000, and 3000~s for Abell~160, Abell~189, and Abell~397 respectively.
Atmospheric seeing quality was between 1.3 and 2.5~arcsec most of the time,
therefore the observed cE galaxies remained spatially unresolved.

The calibrations obtained during the observing run included: (i) bias
frames, (ii) arc line spectra of He-Ne-Ar, (iii) internal spectral flat
fields, (iv) observations of spectrophotometric standard stars, (v) high
signal-to-noise twilight spectra obtained in each configuration of the
multi-slit block.

We reduced the data using the {\sc itt idl} software package. The
primary data reduction steps comprising bias subtraction, flat fielding,
Laplacian filtering \cite{vanDokkum01} for
removing cosmic-ray hits were
applied to all science and calibration frames. We then traced the slitlet
positions using flat-field spectra, estimated the cross-dispersion geometrical
distortions and corrected for them using two-dimensional polynomial
warping. Then, for every individual slitlet we built the wavelength solution
by identifying arc lines and fitting their positions with 
two-dimensional polynomial of the 3rd order in both dimensions, along
and across dispersion, and linearized the spectra. The obtained wavelength
solutions had fitting residuals of about 0.1~\AA~RMS mostly due to the
statistical errors in determined arc line positions. These positions were taken
into account later in the data analysis step. Error frames were computed
using the photon statistics and processed through the same reduction steps
as the data.

In order to precisely measure stellar velocity dispersions, it is essential
to take into account systematic errors in the obtained wavelength solutions
and to get precise information about the spectral resolution variations along
and across dispersion. To do this, we fitted the high-resolution ($R=10000$)
solar spectrum against the linearized twilight spectra in five wavelength
segments overlapping by 20~per~cent, covering the spectral range of the
SCORPIO setup using the penalized pixel fitting procedure\cite{CE04}.

We then extracted one-dimensional spectra of the cE candidate galaxies from
the corresponding slitlets and subtracted the night sky emission
spectrum reconstructed from regions of the same slitlets not including the
target galaxies. In the case of Abell~397, we took into account the
contamination of the ACO~397~J025631.27+155531.2 (A397cE1) spectrum by the
cD galaxy by creating a background spectrum from the part of the slitlet
covering the region of the cD. This was symmetric along the major axis to
that where the A397cE1 galaxy is projected. We analysed independently all
the one-dimensional spectra obtained.

We used the {\sc NBursts} full spectral fitting technique \cite{CPSK07} with
high-resolution ($R=10,000$) {\sc pegase.hr} \cite{LeBorgne+04} simple
stellar population (SSP) models to extract kinematics and stellar populations
from our spectra. The fitting procedure comprised the following steps: (i) a
grid of SSP spectra with a fixed set of ages (spaced nearly logarithmically
from 20~Myr to 18~Gyr) and metallicities (from $-$2.0 to $+$0.5~dex) was
convolved with the wavelength-dependent instrumental response of SCORPIO
obtained from the analysis of twilight frames\cite{CPSA07};
(ii) a non-linear least square fitting against the observed spectrum was done
for a template picked from the pre-convolved SSP grid interpolating on age
($\log t$), and metallicity ($Z$). It was broadened according to the line-of-sight velocity
distribution (LOSVD) parametrized by $v$ and $\sigma$ and
multiplied pixel by pixel with an $n^{\rm{th}}$ order Legendre polynomial
(multiplicative continuum). This resulted in $n + 5$ parameters to be determined
(we used $n=10$ for our SCORPIO data). BTA spectra with their
best-fitting templates are shown in Fig~\ref{specBTA}.

It was shown \cite{Chilingarian09}, that more optimal usage of information
contained in absorption line spectra using full spectral fitting techniques,
improved the precision of age and metallicity estimates by a factor of a
few, as compared to individual absorption line strength indices. This
explains the low statistical uncertainties on the stellar population
parameters presented in Tables~S1--S2. It was also demonstrated
\cite{CPSA07,Chilingarian+08,CCB08} that velocity dispersion measurements
remain precise down to at least 1/2 of the instrumental spectral resolution.
It was also shown that stellar population measurements remain unbiased for
the non-solar $\alpha$/Fe element abundance ratio.  However, as any other
stellar population analysis technique, {\sc NBursts} produces
model-dependent results which may be systematically different if one uses
different stellar population models. All data points in Fig.~1, except those
representing Coma compact stellar systems \cite{Price+09} were obtained
using the {\sc NBursts} spectral fitting technique with the same stellar
population models. It is therefore correct to compare them. We did not
have the original spectra for the Coma compact stellar systems, therefore
the published age and metallicity values \cite{Price+09} were used.

We applied the same fitting technique to available SDSS DR7 spectra of 6
candidate cE galaxies using the wavelength-dependent spectral resolution
information provided for every individual dataset by the SDSS archive. We
similarly fitted 140 SDSS DR7 spectra of early-type galaxies in the central
region of the Coma cluster (Abell~1656) in order to build the
Faber--Jackson\cite{FJ76} and luminosity--metallicity relations for
intermediate-luminosity and giant galaxies displayed in Fig.~1.

\section*{Results}

\begin{figure}
\includegraphics[width=0.32\hsize]{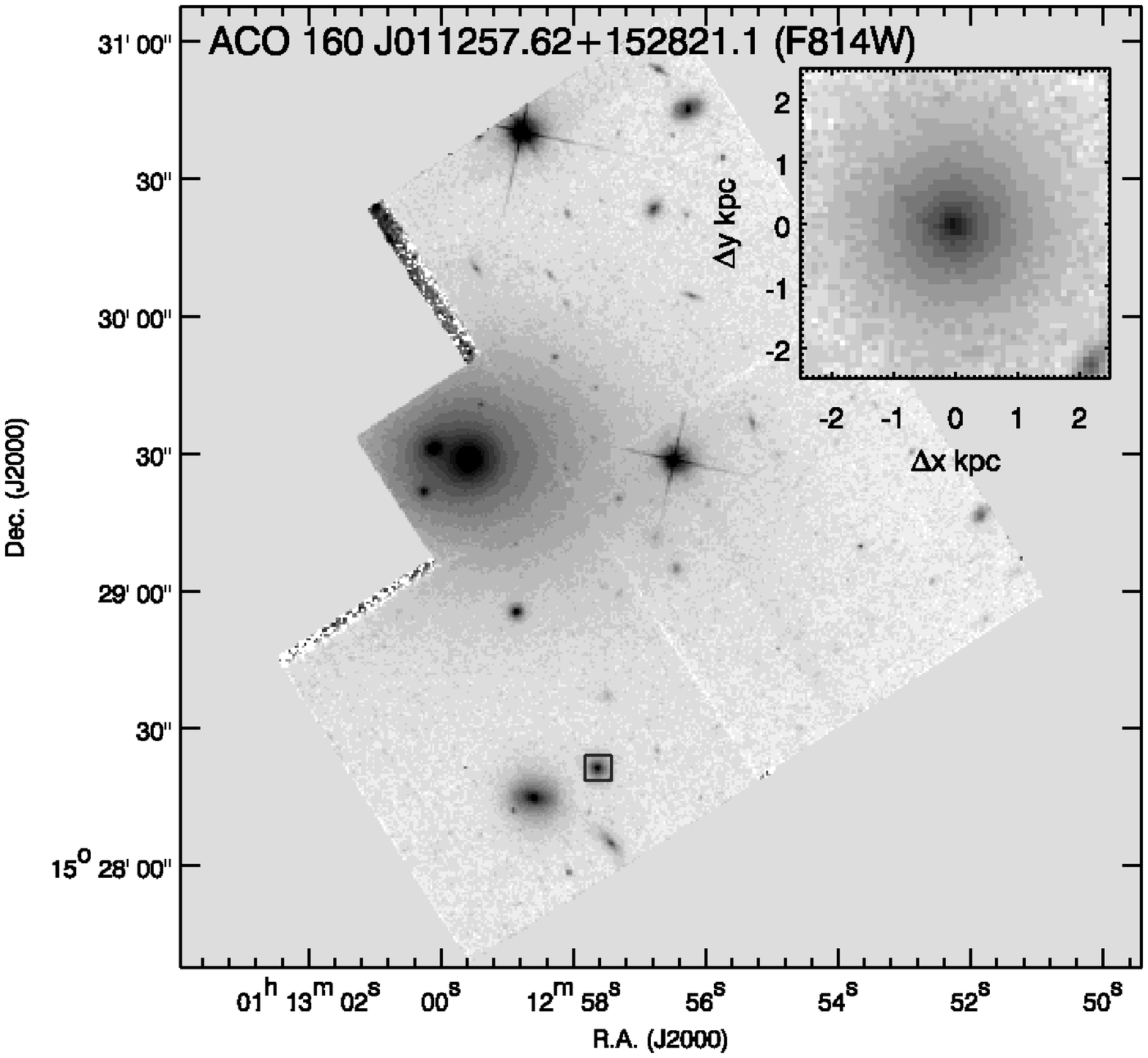}
\includegraphics[width=0.32\hsize]{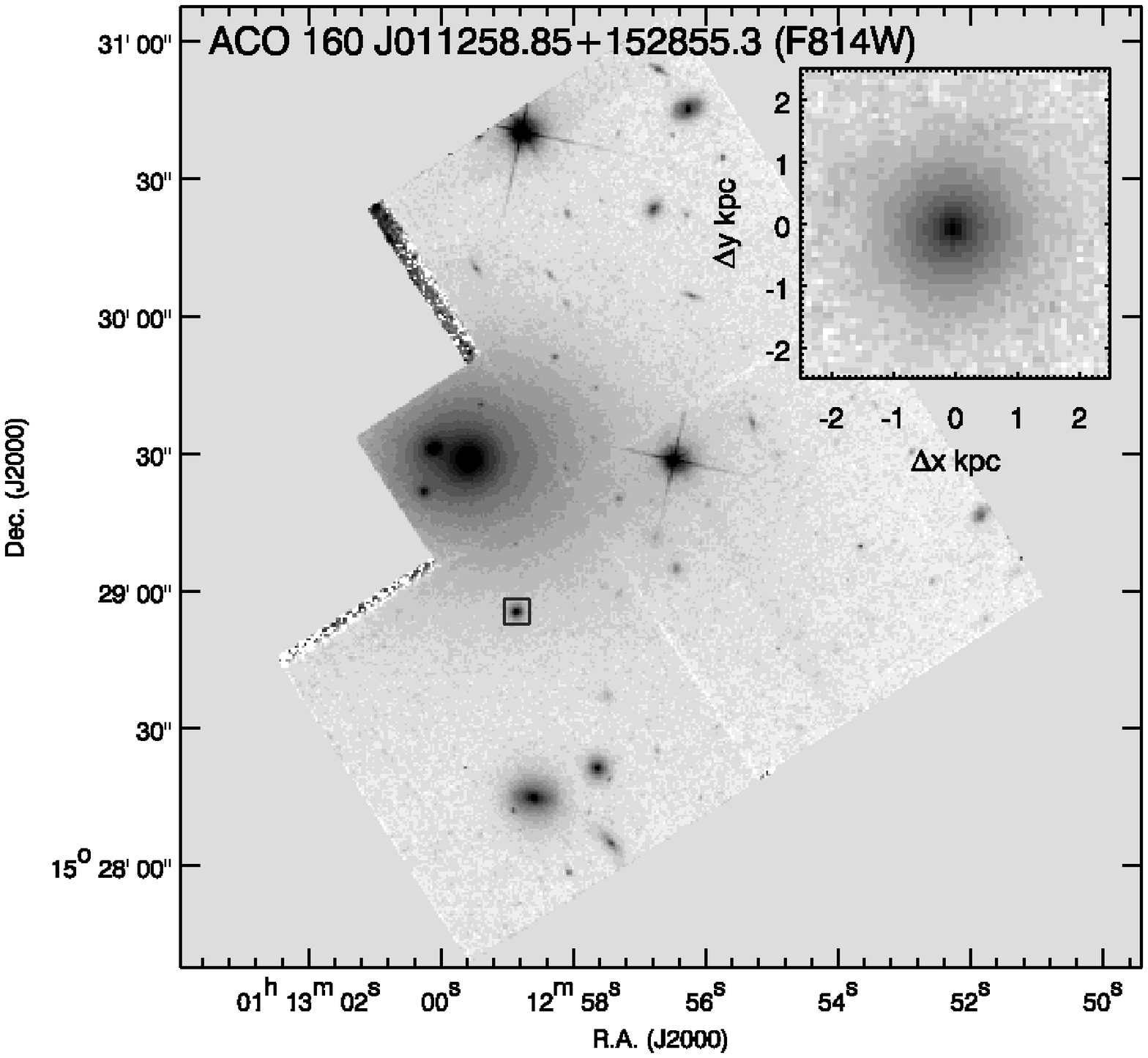}
\includegraphics[width=0.32\hsize]{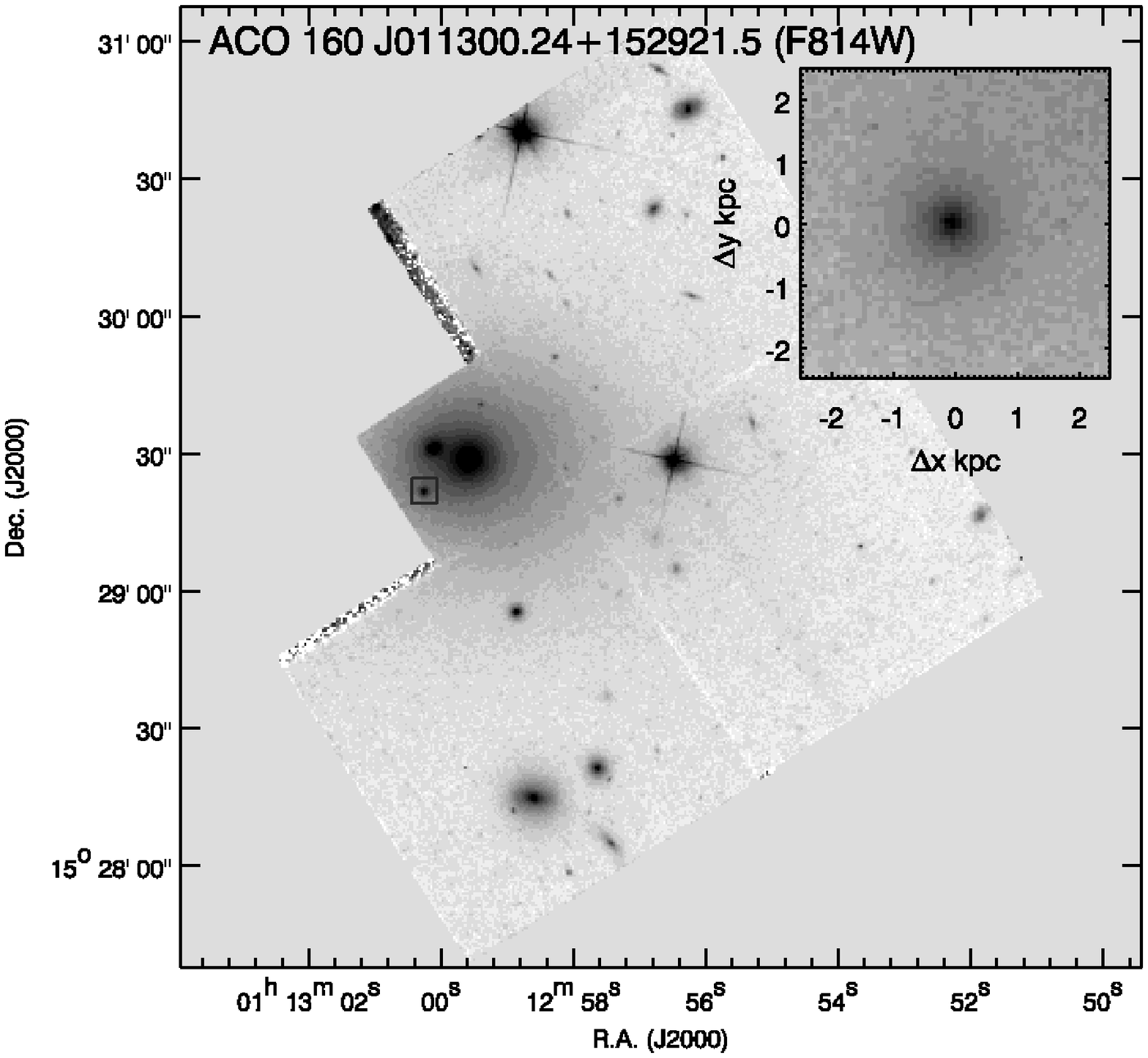}\\
\includegraphics[width=0.32\hsize]{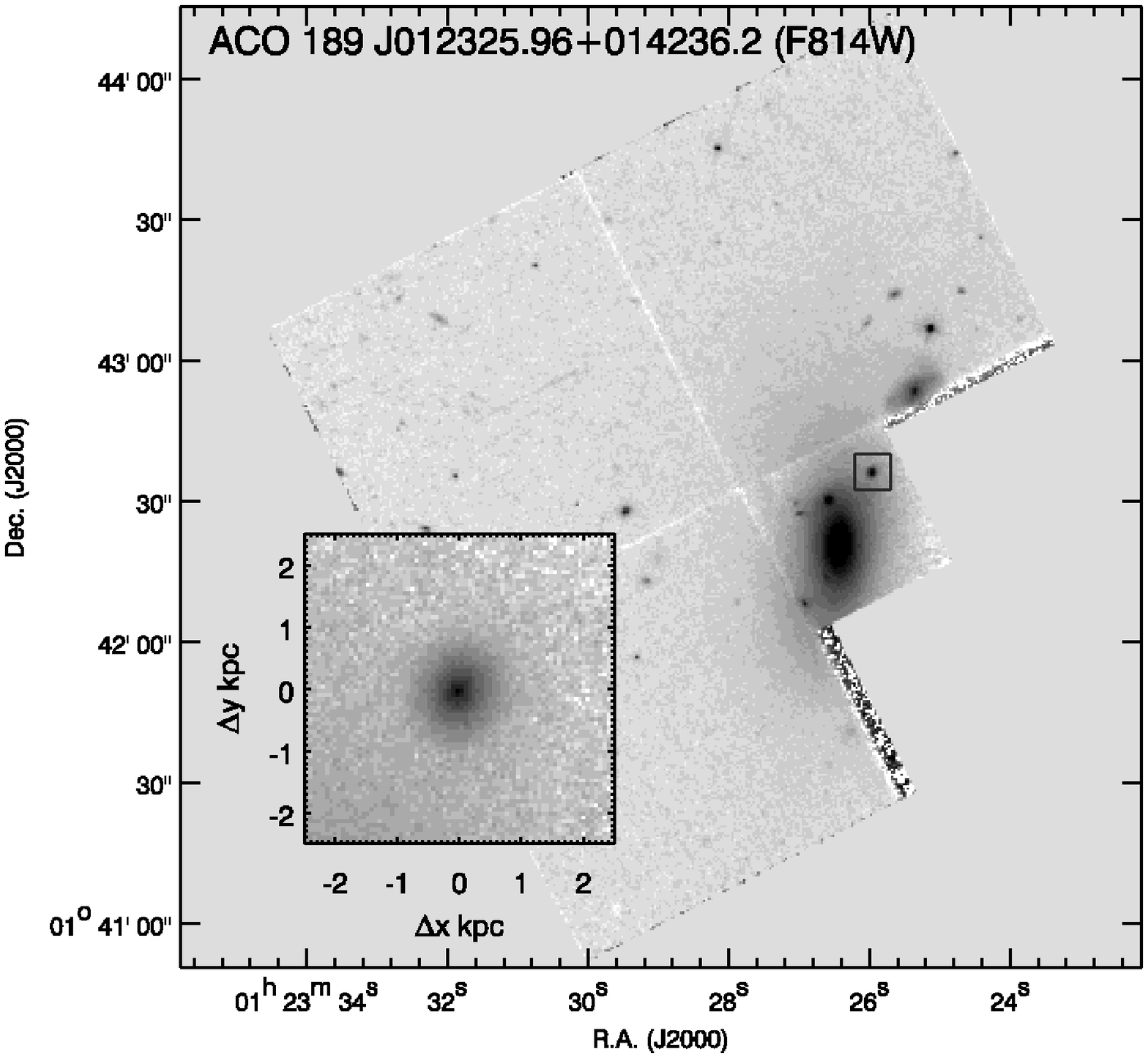}
\includegraphics[width=0.32\hsize]{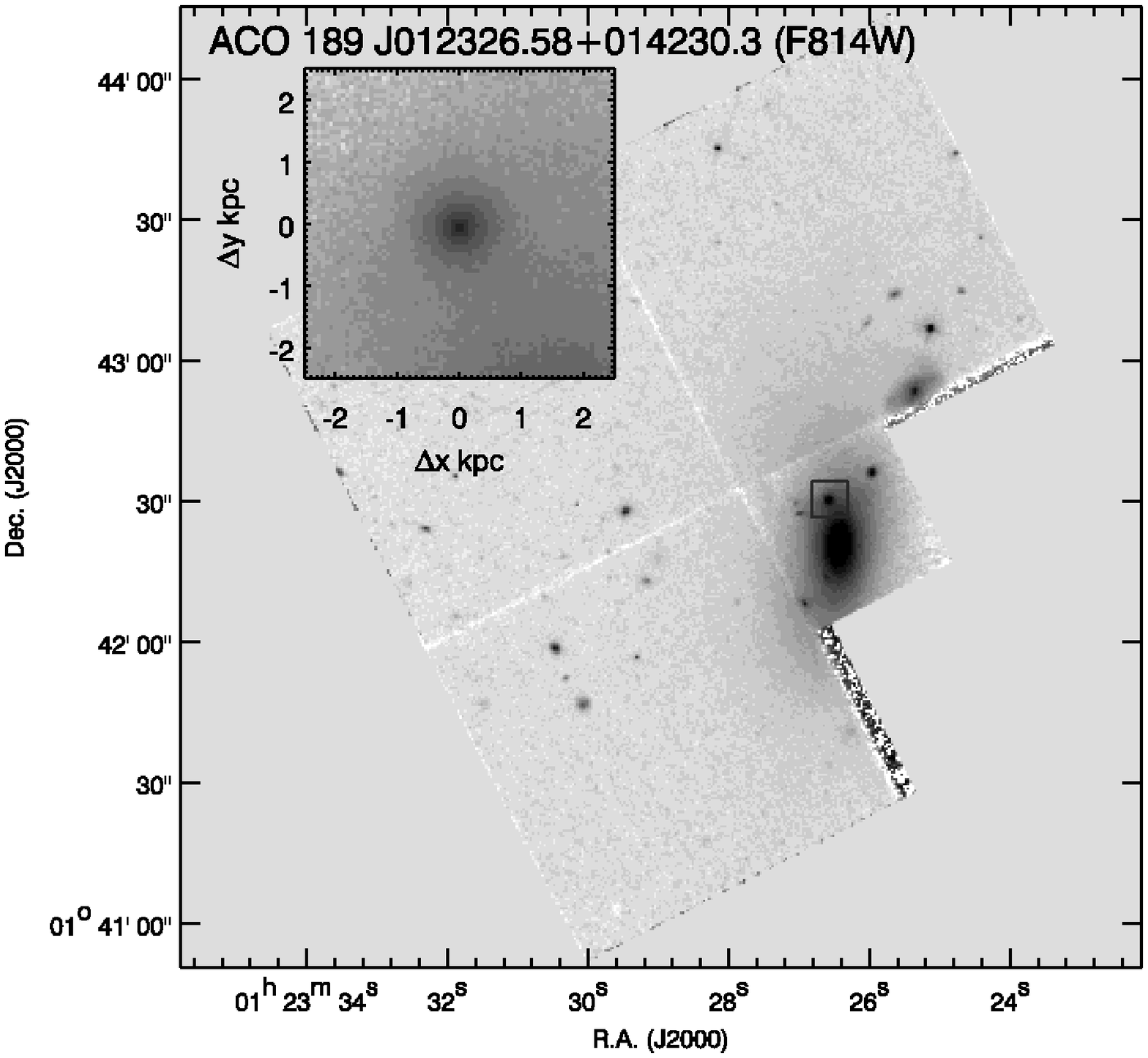}
\includegraphics[width=0.32\hsize]{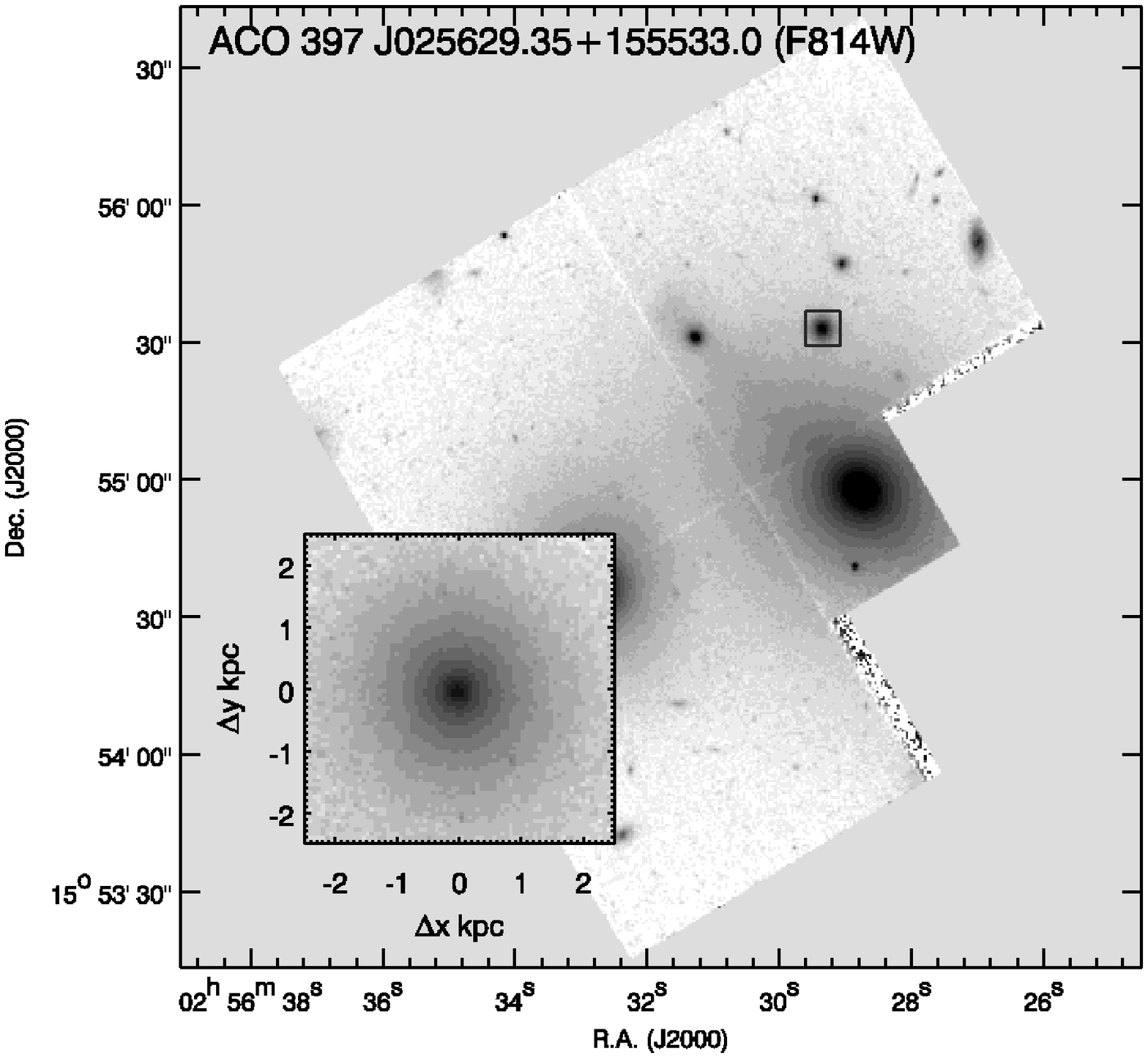}\\
\includegraphics[width=0.32\hsize]{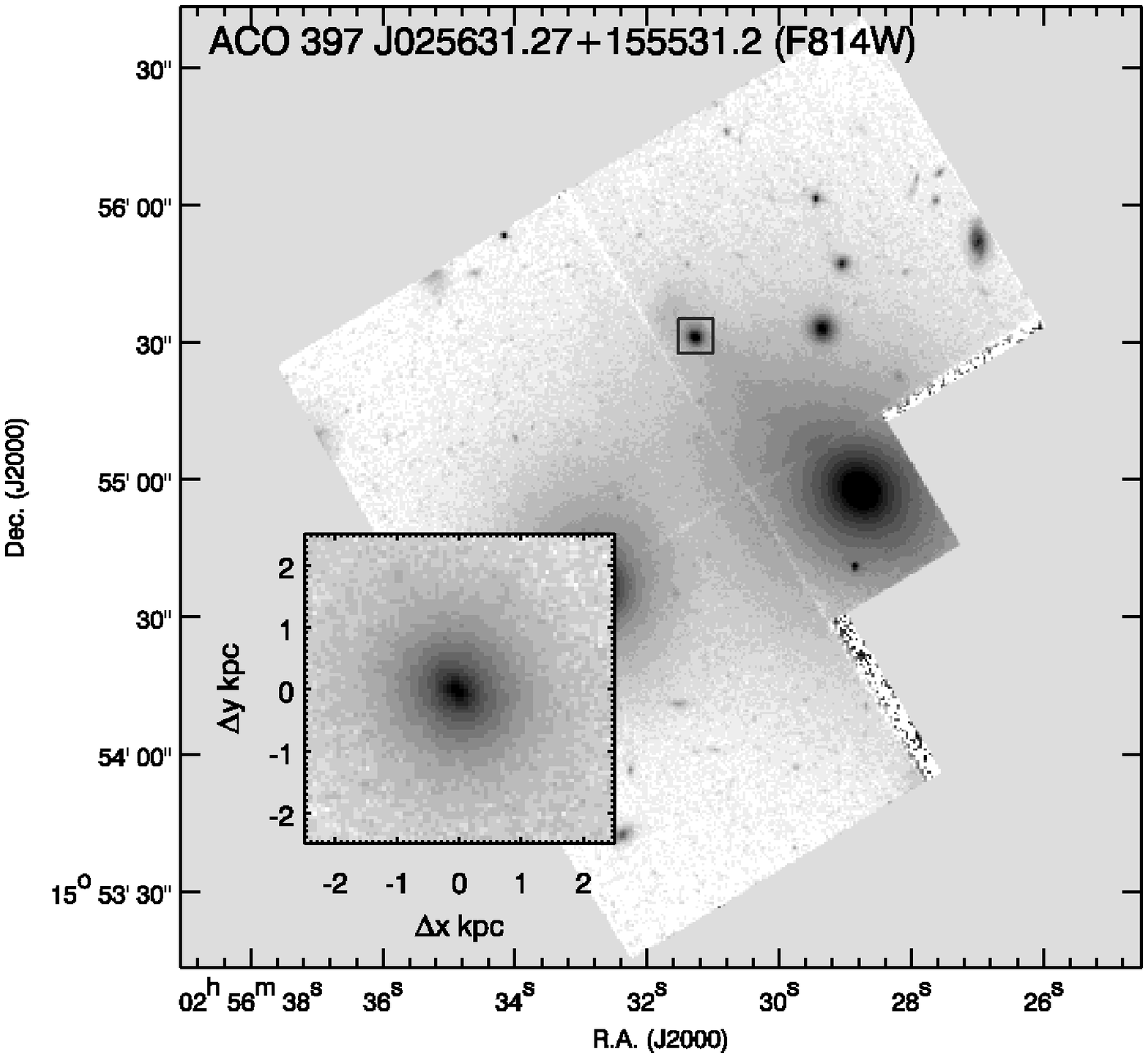}
\includegraphics[width=0.32\hsize]{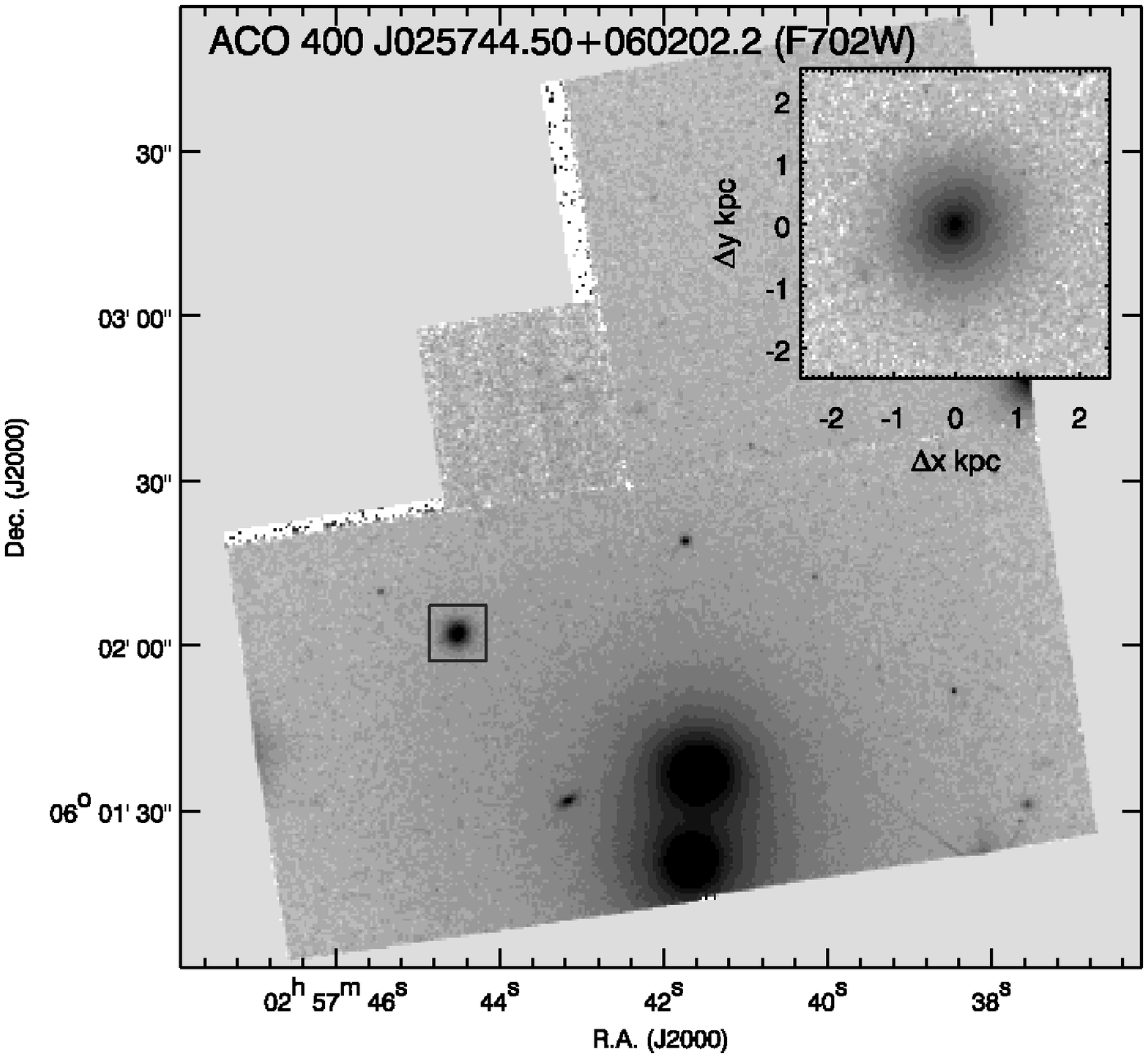}
\includegraphics[width=0.32\hsize]{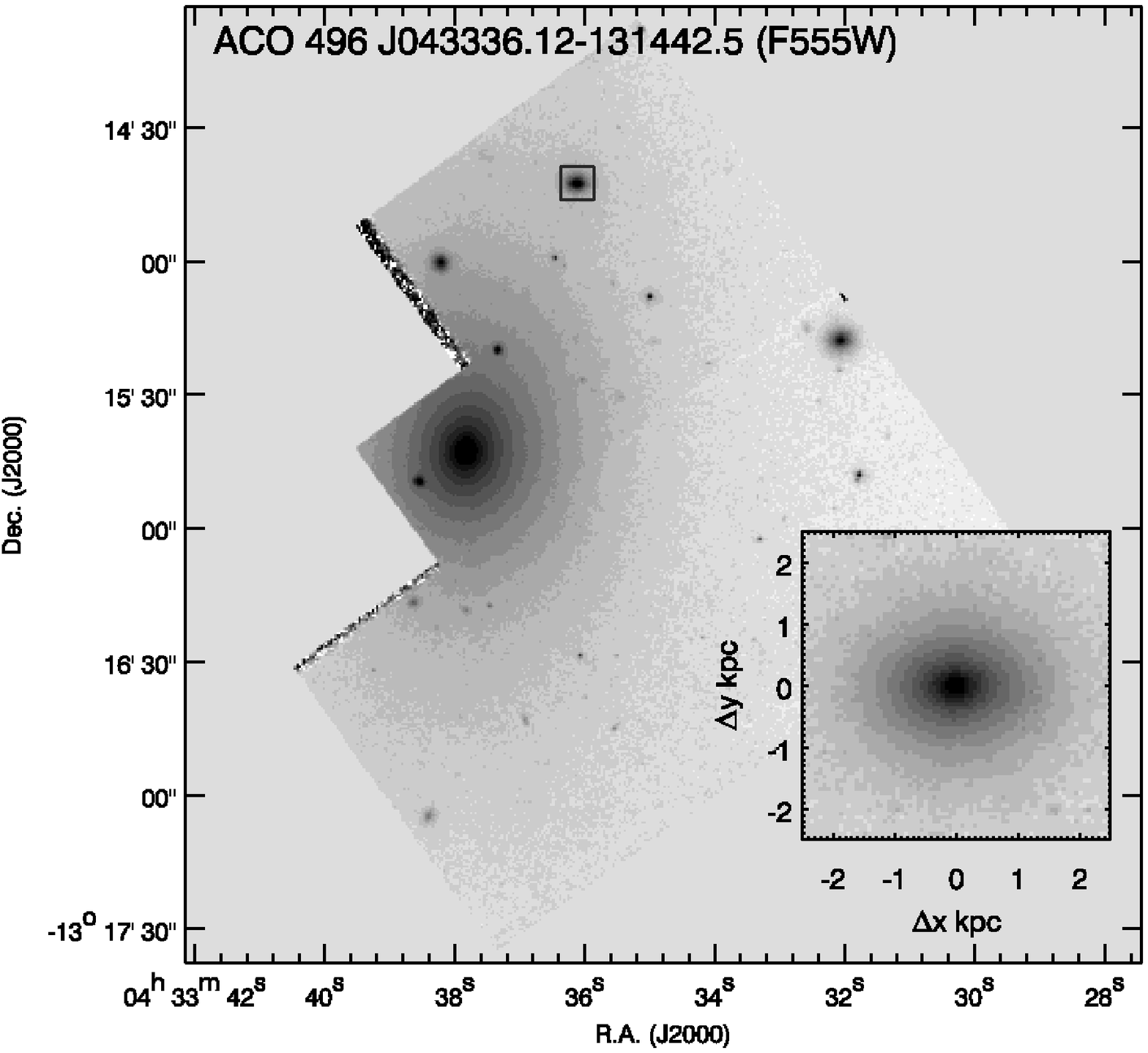}\\
\caption{HST WFPC2 images of cE galaxies. For every galaxy we
show the entire WFPC2 image and an inner panel displaying a $5\times5$~kpc
region centered on the cE galaxy. Cluster identification, IAU recommended
designation of the cE galaxy and the photometric bandpass are
indicated.\label{figmaps}}
\end{figure}

In Fig.~\ref{figmaps} we provide HST WFPC2 images for all confirmed cE
galaxies. For each galaxy we display: (a) the full WFPC2 field of view often
containing a bright cluster galaxy or a cD galaxy and (b) a zoomed cutout
featuring the cE galaxy corresponding to a spatial size of 5$\times$5~kpc at
the cluster distance.

In Tables~\ref{tab1}--\ref{tab2} we present the photometric, kinematic, and
stellar population properties of cE galaxies with confirmed cluster
membership. The columns in Table~\ref{tab1} provide (from left to right):
record numbers, cluster number in the Abell catalogue\cite{Abell58}, cluster
heliocentric radial velocities from NED, IAU-recommended names, absolute
$B$-band magnitudes, half-light radii, mean surface brightnesses within
$r_e$ and ellipticities of the cE galaxies. In Table~\ref{tab2} for each cE
galaxy we provide: projected distances from the cD galaxies, heliocentric
radial velocities, stellar velocity dispersions, luminosity weighted ages
and metallicities, and sources of redshift and photometric data (absolute
magnitude and effective surface brightness). The half-light radii $r_e$, and
ellipticities in Table~\ref{tab1} are always measured from HST WFPC2
imaging, using the SExtractor software\cite{BA96}, while the sources of
total magnitudes vary. We give preference to the ground-based data (SDSS,
CADC MegaPipe Stacks) if their photometric bandpass is closer to the
$B$-band than the HST bandpass. We apply colour transformations\cite{FSI95}
assuming an elliptical galaxy spectrum, in order to convert measurements
into the $B$ band. Corresponding mean surface brightnesses $\langle \mu
\rangle_{e, B}$ are computed from the HST WFPC2 measured half-light radii
and adopted $B$-band magnitudes; flat colour profiles are assumed. All
magnitude and surface brightness values are corrected for Galactic
extinction\cite{SFD98} and $K$-corrections. The surface brightness values
are additionally corrected for cosmological dimming.

In addition to the galaxies presented in
Tables~\ref{tab1},\ref{tab2}, our workflow re-detected two objects in the
Abell~496 cluster which were referenced before as A496cE and
A496g1\cite{Chilingarian+07}. They are shown as known cE galaxies in
Fig.~1, but we have intentionally removed them from
Tables~\ref{tab1},\ref{tab2}.

\section*{$N$-body simulations of tidal stripping}

Spatial resolution of cosmological simulations is insufficient to resolve
the internal structure of intermediate luminosity galaxies in detail. A
number of studies addressed dynamical and morphological evolution of disc
galaxies in clusters from simulations
\cite{Moore+96,AMB99,QMB00,Mastropietro+05}; however, none of them 
analyzed strong tidal stripping and tidal disruption. The nature of
Messier~32, the prototypical cE galaxy, was explained by tidal threshing
\cite{BCDG01} but it was not shown how this process would act on progenitors
having larger masses in a cluster environment. This motivated us to perform
32 dedicated intermediate-resolution ($\sim 100$~pc) $N$-body simulations of
interactions of a disc galaxy with a galaxy cluster potential using the {\sc
gadget-2} treecode\cite{Springel05}. We only considered the gravitational
forces, neglecting the ram pressure stripping effects.


The disc galaxy is based on the ``classical model''\cite{RPCB09}. It is made
from a bulge ($0.42\,\times 10^{11}\,\rm{M_\odot}$), a
Miyamoto--Nagai\cite{MN75} gaseous disc ($0.3\,\times
10^{11}\,\rm{M_\odot}$), an exponential stellar disc ($1.04\,\times
10^{11}\,\rm{M_\odot}$) and a Plummer dark matter halo ($8.9\,\times
10^{11}\,\rm{M_\odot}$).  The relative masses of all components are set,
such as the rotation curve being flat up to 40~kpc.  The total mass is $9.6 \times
10^{11} M_{\odot}$. As an initial condition, we took a disc containing a
central bar. The extension of the stellar bar is about 4~kpc with an axial
ratio $a/b = 1.8$. Its rotation period is 220~Myr
($\Omega_p=28$~km~s$^{-1}$~kpc$^{-1}$) placing the inner Lindblad resonance
at 3.5~kpc and the corotation radius at 8.4~kpc.

The galaxy cluster potential model is based on the data of
Messier~87\cite{Emsellem+04}, the Virgo cluster cD. It is modeled by a live
NFW\cite{NFW97} potential represented by $5 \times 10^5$ particles with a
concentration of 7~kpc and a virial mass of $1.4 \times
10^{14}$~M$_{\odot}$. We also took into account the central giant black hole
of $3 \times 10^9$~M$_{\odot}$, represented by a Plummer sphere with a
core of 100~pc, which is similar to the resolution of our simulations.

We performed 32 simulations varying the orbital angular momentum
(initial tangential velocities of 650, 180, 120, and 50~km~s$^{-1}$
corresponding to initial ellipticities of 0.04, 0.79, 0.86, and 0.94)
and the disc inclination with respect to the orbital plane (0, $\pm$45,
$\pm$90, $\pm$135, and 180~deg). Each simulation was followed for 2~Gyr.
Initially the disc galaxy was located 50~kpc from the centre of the cD
potential.

The surface density maps presented in Fig.~3 are computed from the stellar
components by adding the cD stellar density profile. It was modelled by
fitting the light distribution of Messier~87\cite{KFCB09} with a
Hernquist\cite{Hernquist90} profile and assuming constant mass-to-light
ratio along the radius. The outermost contour corresponds to the $B$-band
surface brightness value $\mu_B = 28$~mag~arcsec$^{-2}$. The contours are
shown with a step of 2~mag~arcsec$^{-2}$.

In Fig.~1 we present the evolution of structural and dynamical properties of
a stripped galaxy. We took initial ages of 4 and 10~Gyr for the bulge and disc
components. We computed at every timestep the corresponding mass-to-light
ratios from the {\sc pegase.2} stellar population models \cite{FR97} in
order to convert surface density into $B$-band surface brightness. The
resulting maps were analyzed by the SExtractor software in the same way as the
HST images, which we used to search for cEs in galaxy clusters. The black
path displays the initial conditions and snapshots at 50, 250, 500, 750,
1200, 1500, and 1950~Myr.

Our simulations demonstrate that the large-scale stellar discs get heavily
stripped even on the quasi-circular orbit ($e=0.04$) 600--800~Myr after
the start of the simulation. About 50~per~cent of the initial stellar mass
of the galaxy is stripped. However, since only the disc, having lower
mass-to-light ratio than the bulge is affected, the total predicted $B$-band
luminosity decreases by more than a factor of 3.

The progenitor's bar remains weakly affected and looks structurally like a
high surface brightness flattened elliptical galaxy resembling
ACO496~J043336.12-131442.5. In case of high disc inclination to the initial
orbital plane, at certain moments in time, the remnant's major axis becomes
nearly parallel to the line of sight, creating the effect of very high
surface brightness round galaxies. On such a quasi-circular orbit, the
stripped remnant may remain for quite a long time without further
significant mass loss. We note that the tidal stripping scenario is
compatible with the low-surface brightness outer exponential discs detected
in Messier~32\cite{Graham02} and A496cE\cite{Chilingarian+07}.

Interactions on radial orbits ($e=0.86 \dots 0.94$) result in more efficient
stripping with a stellar mass loss reaching 85--90~per~cent 600--700~Myr
after the start of the simulation. The first 70--75~per~cent are already
lost at $t=$250--300~Myr. The outer isophotes of the remnant may become
rounder due to the interplay between tidal compression in the tangential
directions and rotation of the barred galaxy. Correspondingly, the remnant's
mean surface density changes periodically, being connected to the orbital
motion. 
However, in all cases of radial orbits, the remnants are already
accreted by the cD galaxy at $t=$1.0--1.2~Gyr.

Thus, we have two alternatives: (i) galaxies on quasi-circular orbits may
remain for a long time without suffering significant structural changes
after the initial stripping of their large-scale discs removing a significant
fraction of the stellar mass; (ii) galaxies infalling on radial orbits lose
mass more efficiently and over a shorter time. However, they are completely
accreted by the cD on a timescale of about 1~Gyr, so one needs a supply of
infalling galaxies on to the cluster centre in order to form cEs. 
Scaling the masses of progenitors, we can create compact galaxies ranging in
luminosities. Given the dependence of the galaxy profile concentration on
the luminosity\cite{Ferrarese+06,GG03}, one can expect lower central
surface densities for tidally stripped galaxies originating from lower-mass
progenitors.

A large fraction of the stripped material is not accreted by the cD galaxy
and contributes to the diffuse intracluster light. Analysing its properties
will help us to estimate the frequency of tidal stripping events and
understand their importance as a mechanism of galaxy evolution in dense
environments.


\renewcommand\refname{Supporting References and Notes}

%
%

\clearpage

\begin{table}
\caption{Photometric properties of confirmed cE galaxies and their host
clusters\label{tab1}}
\begin{tabular}{llllcccc}
\hline
\hline
N & Cluster & $cz_{\mbox{clus}}$ & IAU Name & $M_B$ & $r_e$ & 
   $\langle \mu \rangle_{e, B}$ & $\epsilon$ \\
  & & km~s$^{-1}$ & & mag & kpc & mag~arcsec$^{-2}$ & \\
\hline
  1 & ACO~0160 & $13401$ & $J011257.62+152821.1$ & $-17.08$ & $ 0.56$ & $20.24$ & $ 0.02$\\
  2 & ACO~0160 & $13401$ & $J011258.85+152855.3$ & $-16.87$ & $ 0.43$ & $19.87$ & $ 0.03$\\
  3 & ACO~0160 & $13401$ & $J011300.24+152921.5$ & $-16.26$ & $ 0.33$ & $19.89$ & $ 0.01$\\
  4 & ACO~0189 & $ 9833$ & $J012325.96+014236.2$ & $-15.84$ & $ 0.31$ & $20.16$ & $ 0.14$\\
  5 & ACO~0189 & $ 9833$ & $J012326.58+014230.3$ & $-15.78$ & $ 0.44$ & $21.01$ & $ 0.10$\\
  6 & ACO~0397 & $ 9803$ & $J025629.35+155533.0$ & $-17.63$ & $ 0.62$ & $19.91$ & $ 0.09$\\
  7 & ACO~0397 & $ 9803$ & $J025631.27+155531.2$ & $-17.58$ & $ 0.39$ & $18.96$ & $ 0.12$\\
  8 & ACO~0400 & $ 7315$ & $J025744.50+060202.2$ & $-17.16$ & $ 0.42$ & $19.54$ & $ 0.08$\\
  9 & ACO~0496 & $ 9863$ & $J043336.12-131442.5$ & $-18.16$ & $ 0.37$ & $18.23$ & $ 0.23$\\
 10 & ACO~0496 & $ 9863$ & $J043338.55-131549.5$ & $-17.18$ & $ 0.32$ & $18.87$ & $ 0.14$\\
 11 & ACO~0779 & $ 6742$ & $J091947.87+334604.8$ & $-16.83$ & $ 0.32$ & $19.23$ & $ 0.03$\\
 12 & ACO~1177 & $ 9473$ & $J110947.06+214648.6$ & $-17.43$ & $ 0.49$ & $19.60$ & $ 0.25$\\
 13 & ACO~3526 & $ 3418$ & $J124853.91-411905.8$ & $-16.86$ & $ 0.34$ & $19.36$ & $ 0.03$\\
 14 & ACO~1656 & $ 6925$ & $J125923.41+275510.4$ & $-16.19$ & $ 0.20$ & $18.91$ & $ 0.01$\\
 15 & ACO~1656 & $ 6925$ & $J125942.30+275529.0$ & $-18.39$ & $ 0.47$ & $18.56$ & $ 0.17$\\
 16 & ACO~3558 & $14390$ & $J132758.71-312937.5$ & $-17.39$ & $ 0.44$ & $19.42$ & $ 0.17$\\
 17 & ACO~3562 & $14690$ & $J133340.86-314008.3$ & $-17.06$ & $ 0.30$ & $18.90$ & $ 0.20$\\
 18 & ACO~2040 & $13790$ & $J151250.26+072621.9$ & $-18.13$ & $ 0.65$ & $19.49$ & $ 0.10$\\
 19 & ACO~2052 & $10640$ & $J151641.28+070006.1$ & $-17.04$ & $ 0.35$ & $19.28$ & $ 0.03$\\
 20 & ACO~2634 & $ 9409$ & $J233825.48+270150.1$ & $-17.03$ & $ 0.47$ & $19.87$ & $ 0.07$\\
 21 & ACO~2634 & $ 9409$ & $J233829.31+270225.1$ & $-15.70$ & $ 0.25$ & $19.89$ & $ 0.04$\\
\hline
\hline
\end{tabular}
\end{table}

\begin{table}
\caption{Kinematics and stellar populations of confirmed cE
galaxies\label{tab2}}
\begin{tabular}{lcccccl}
\hline
\hline
N & $d_{proj}$ & $v_r$ & $\sigma$ & $t$ & [Fe/H] & sources of redshifts \\
  & kpc & km~s$^{-1}$ &  km~s$^{-1}$ & Gyr & dex & and photometric data \\
\hline
  1 & $66$ & $ 12645\pm 4$ & $  59\pm6$ & $  8.6\pm  2.8$ & $-0.41\pm0.05$ & $z$:BTA, ph:SDSS ($g$)\\
  2 & $32$ & $ 12549\pm 5$ & $  70\pm6$ & $ 16.2\pm  6.1$ & $-0.37\pm0.17$ & $z$:BTA, ph:SDSS ($g$)\\
  3 & $11$ & $ 12848\pm 4$ & $  48\pm7$ & $  7.5\pm  2.2$ & $-0.38\pm0.05$ & $z$:BTA, ph:SDSS ($g$)\\
  4 & $11$ & $ 10019\pm 4$ & $  61\pm7$ & $ 17.5\pm  4.1$ & $-0.70\pm0.12$ & $z$:BTA, ph:HST ($F814W$)\\
  5 & $6.6$ & $ 10411\pm 3$ & $  48\pm6$ & $  7.4\pm  2.2$ & $-0.02\pm0.06$ & $z$:BTA, ph:HST ($F814W$)\\
  6 & $24$ & $ 10278\pm 4$ & $  68\pm5$ & $  9.0\pm  1.9$ & $-0.06\pm0.04$ & $z$:BTA, ph:CFHT ($g$)\\
  7 & $33$ & $ 10240\pm 4$ & $  91\pm6$ & $ 11.1\pm  2.6$ & $-0.02\pm0.05$ & $z$:BTA, ph:CFHT ($g$)\\
  8 & $25$ & $  7790\pm30$ & $\dots$ & $\dots$ & $\dots$ & $z$:Vizier, ph:SDSS ($g$)\\  
  9 & $43$ & $  9415\pm31$ & $\dots$ & $\dots$ & $\dots$ & $z$:NED, ph:HST ($F555W$) \\  
 10 & $8.4$ & $  9814\pm21$ & $\dots$ & $\dots$ & $\dots$ & $z$:Vizier, ph:HST ($F555W$) \\ 
 11 & $31$ & $  7142\pm 3$ & $ 110\pm4$ & $ 12.0\pm  1.6$ & $-0.40\pm0.03$ & $z$:SDSS, ph:SDSS ($g$) \\
 12 & $54$ & $  9250\pm 3$ & $  98\pm4$ & $  9.6\pm  1.3$ & $-0.15\pm0.02$ & $z$:SDSS, ph:SDSS ($g$) \\
 13 & $13$ & $  2317\pm20$ & $\dots$ & $\dots$ & $\dots$ & $z$:NED, ph:HST ($F555W$) \\ 
 14 & $14$ & $  6925\pm 5$ & $ 108\pm4$ & $ 11.2\pm  2.1$ & $-0.38\pm0.04$ & $z$:SDSS, ph:SDSS ($g$)\\
 15 & $24$ & $  6916\pm 2$ & $ 177\pm2$ & $ 12.7\pm  0.9$ & $-0.28\pm0.02$ & $z$:SDSS, ph:SDSS ($g$)\\
 16 & $24$ & $ 16668\pm20$ & $\dots$ & $\dots$ & $\dots$ & $z$:NED, ph:HST ($F814W$)\\
 17 & $79$ & $ 14746\pm\dots$ & $ 182\pm9$ & $\dots$ & $\dots$ & $z$:Vizier, ph:Vizier ($B$)\\ 
 18 & $42$ & $ 14551\pm 5$ & $ 100\pm7$ & $ 13.0\pm  4.2$ & $-0.30\pm0.07$ & $z$:SDSS, ph:SDSS ($g$)\\
 19 & $62$ & $ 10350\pm 5$ & $ 117\pm6$ & $ 10.0\pm  2.8$ & $-0.38\pm0.03$ & $z$:SDSS, ph:SDSS ($g$)\\
 20 & $38$ & $  9532\pm80$ & $\dots$ & $\dots$ & $\dots$ & $z$:NED, ph:HST ($F555W$)\\ 
 21 & $20$ & $ 10183\pm80$ & $\dots$ & $\dots$ & $\dots$ & $z$:NED, ph:HST ($F814W$)\\ 
\hline
\hline
\end{tabular}
\end{table}

\clearpage

\begin{figure}
\includegraphics[width=0.32\hsize]{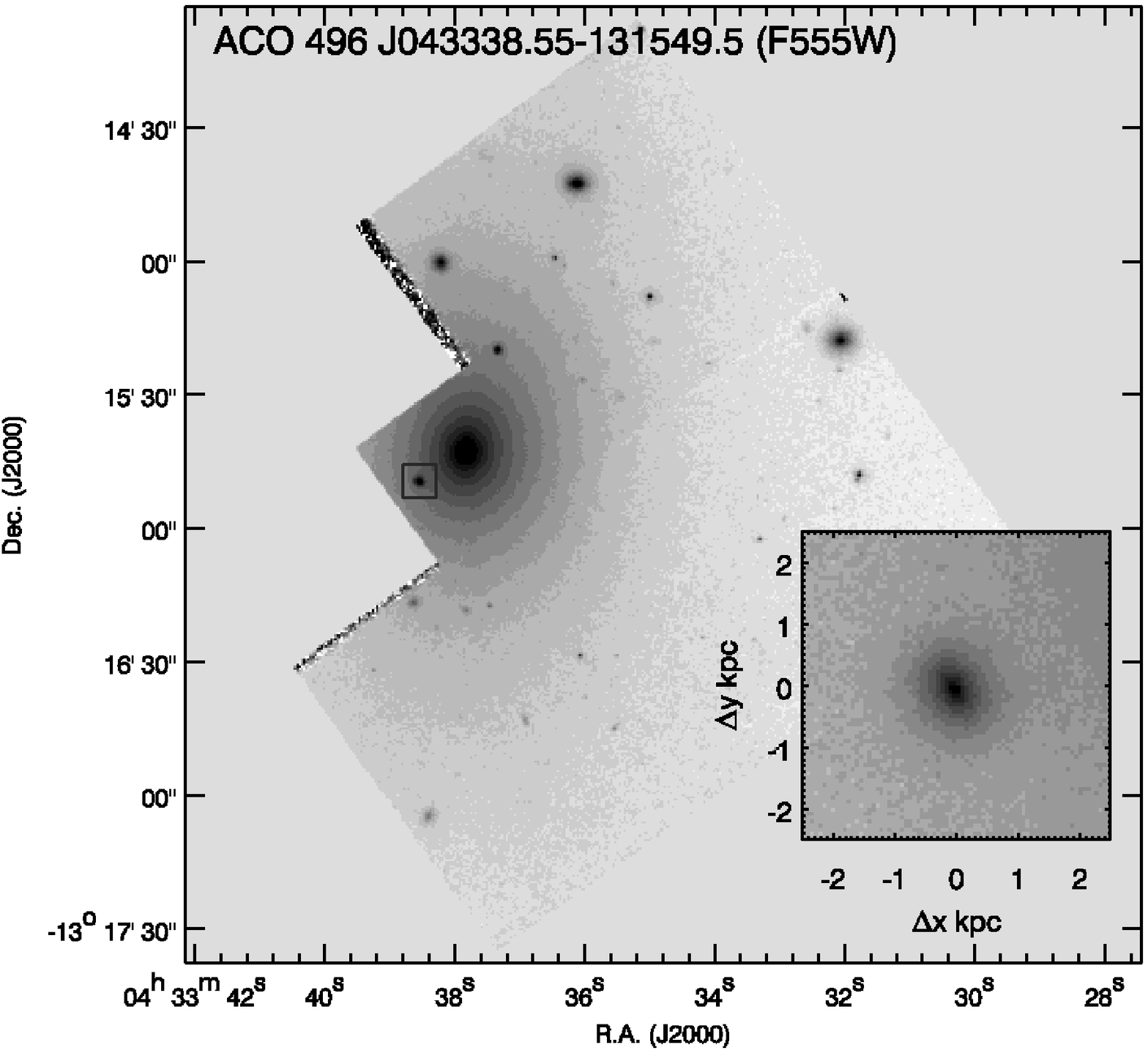}
\includegraphics[width=0.32\hsize]{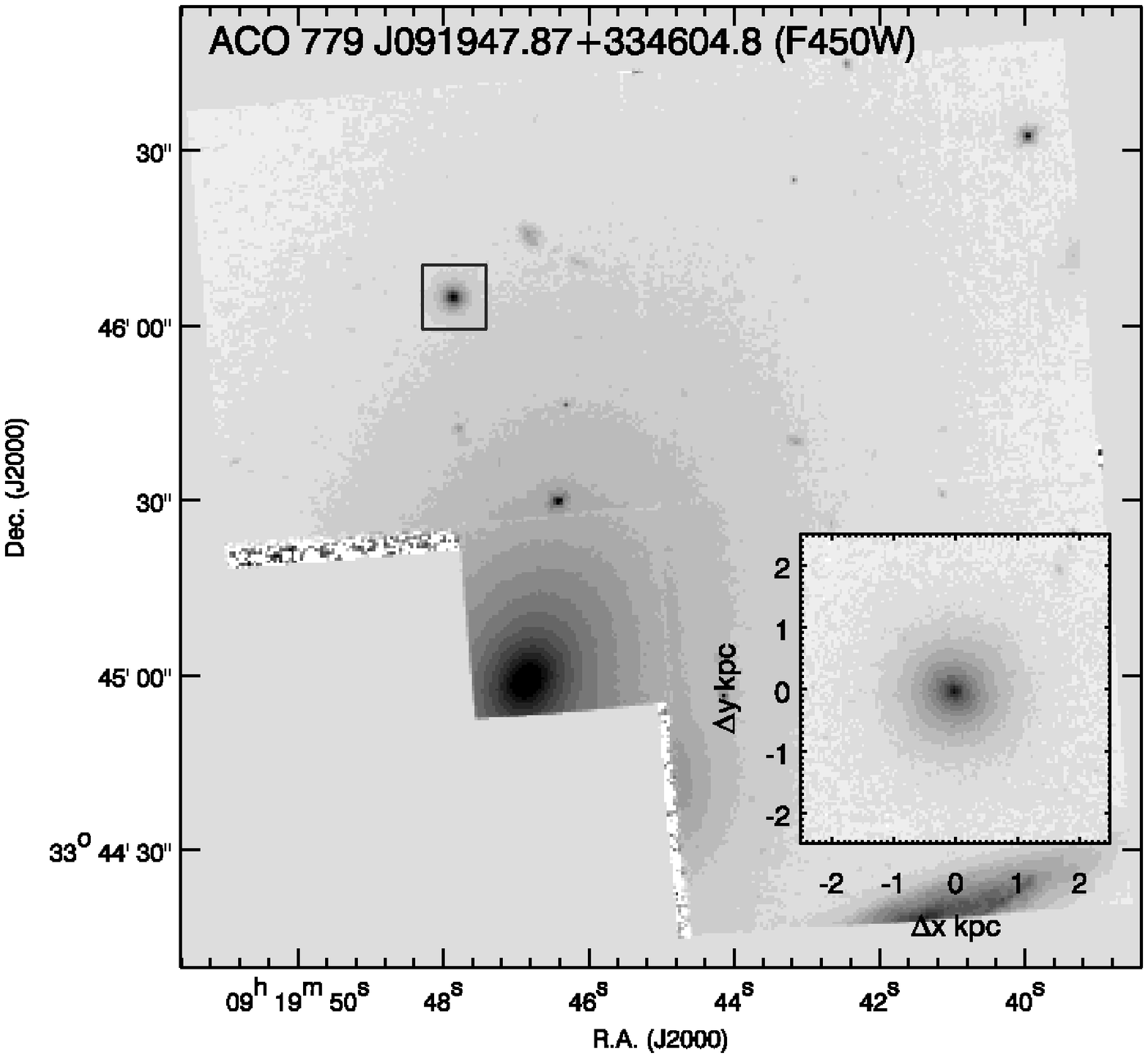}
\includegraphics[width=0.32\hsize]{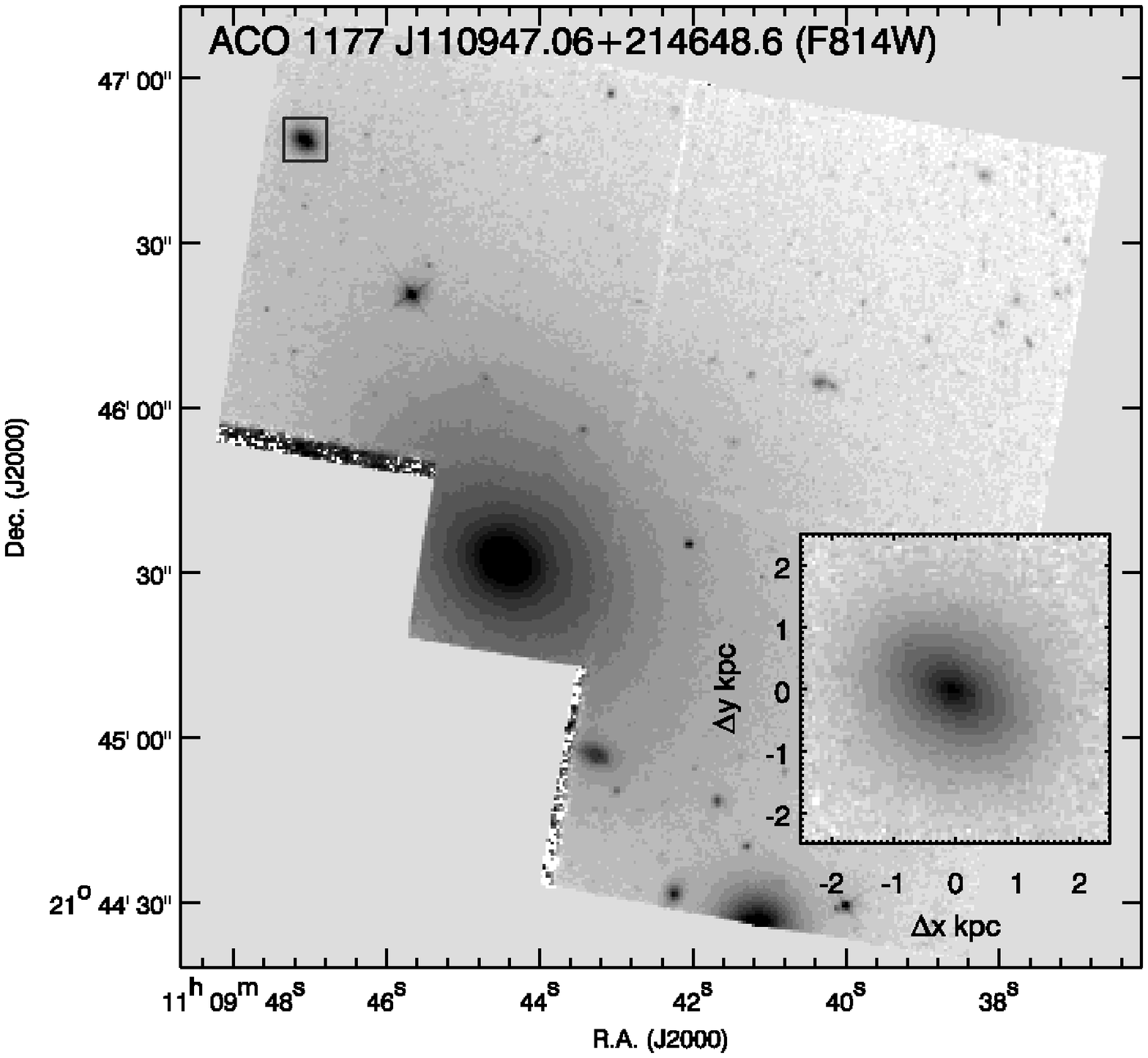}\\
\includegraphics[width=0.32\hsize]{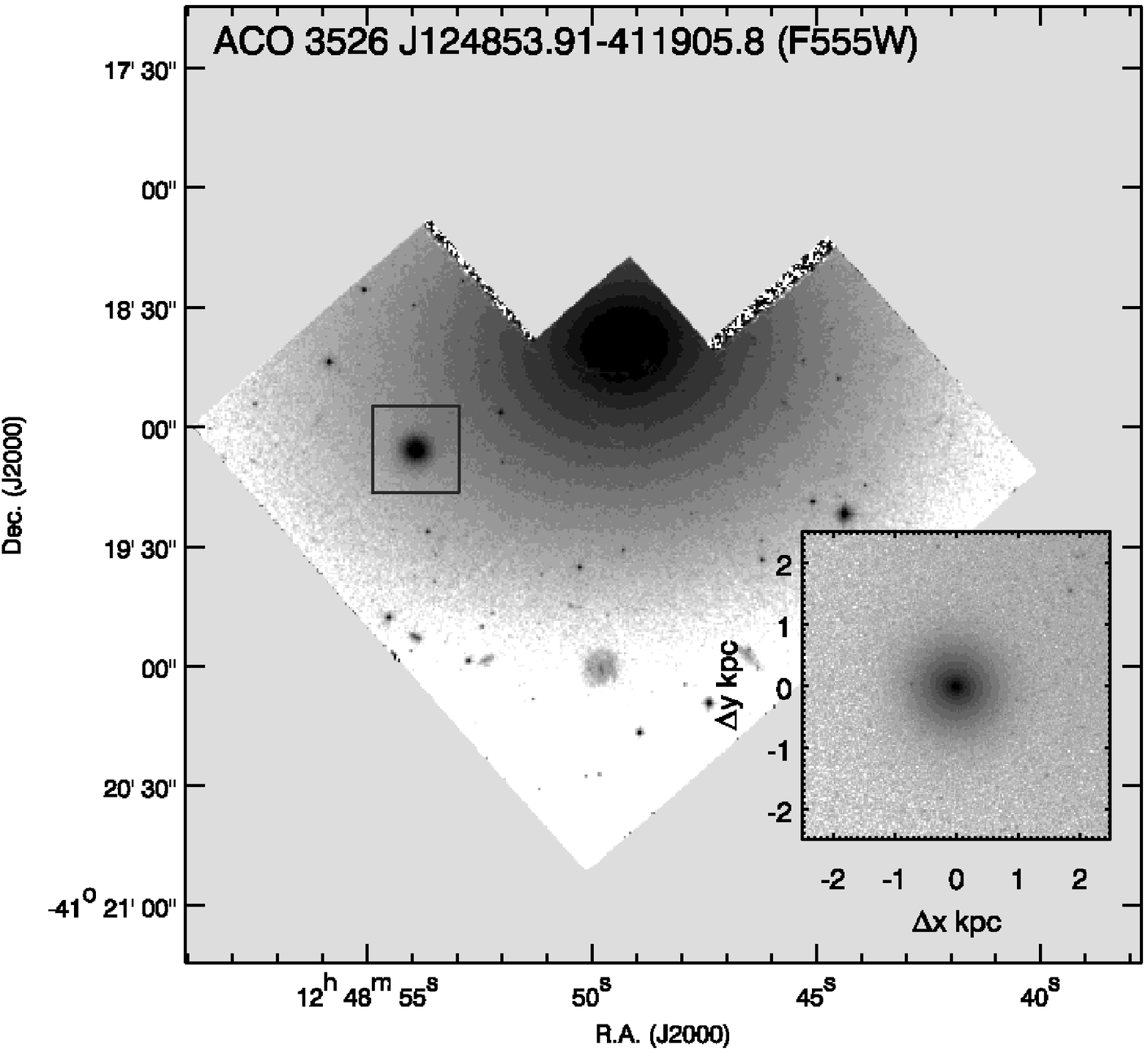}
\includegraphics[width=0.32\hsize]{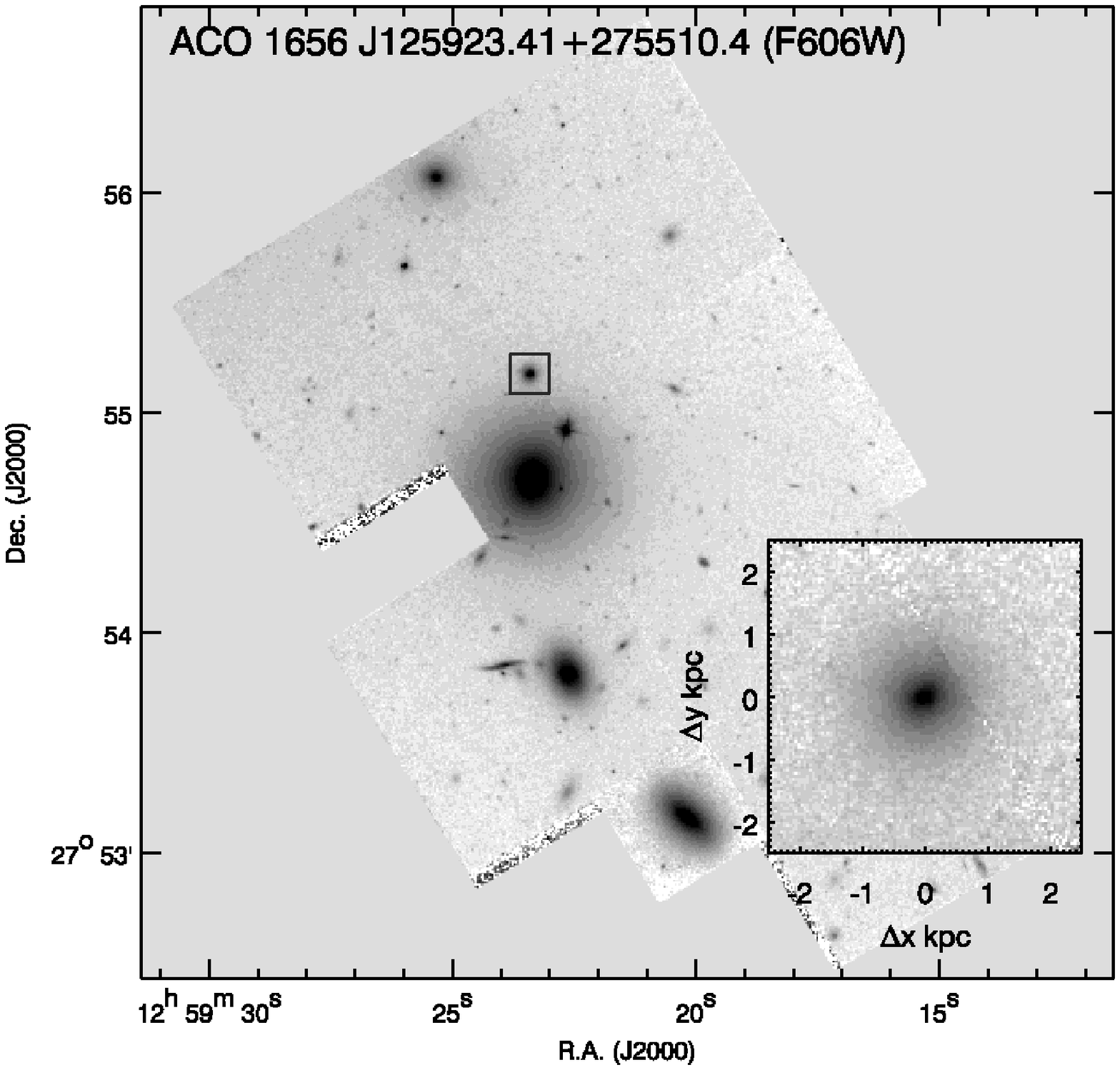}
\includegraphics[width=0.32\hsize]{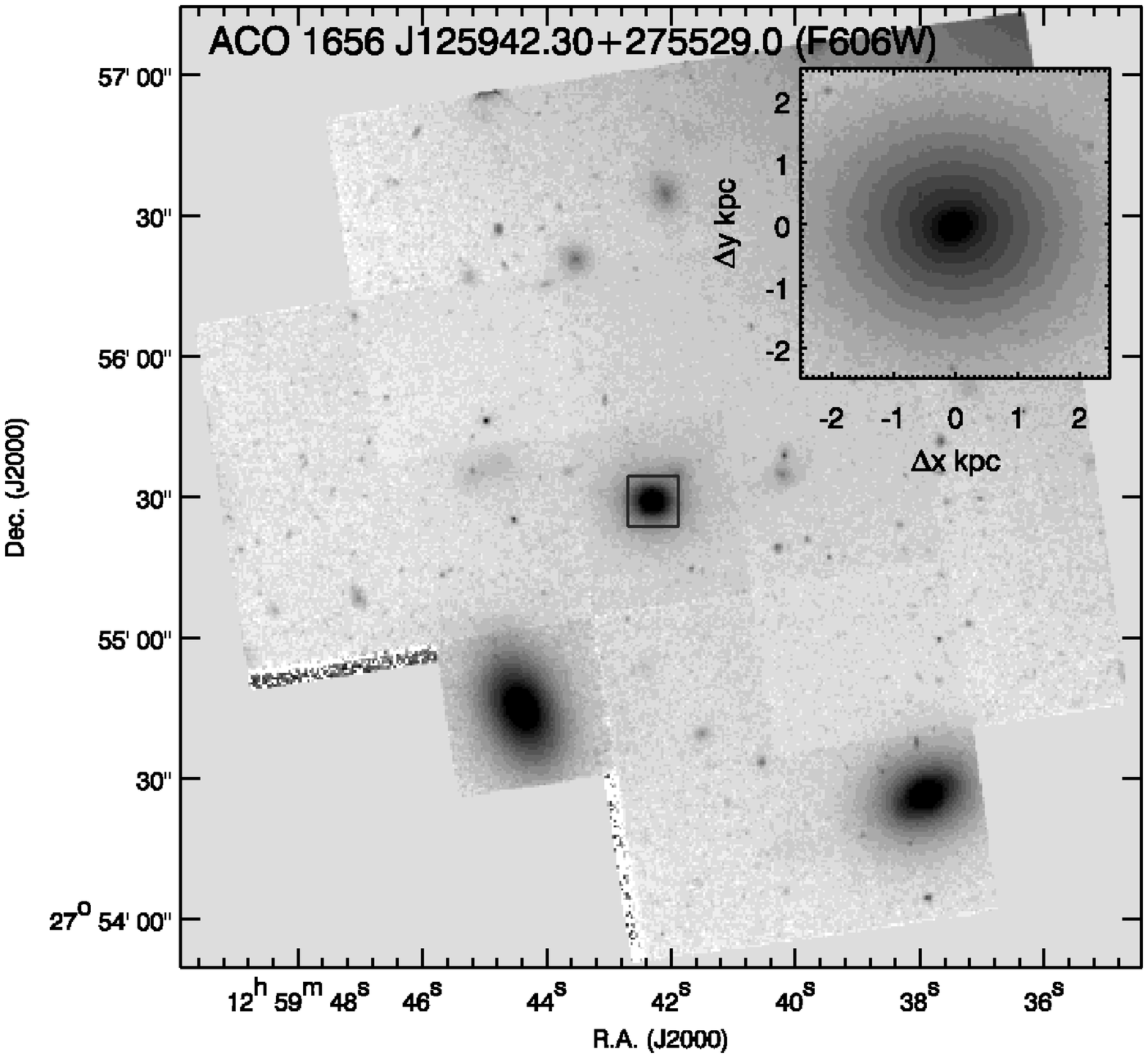}\\
\includegraphics[width=0.32\hsize]{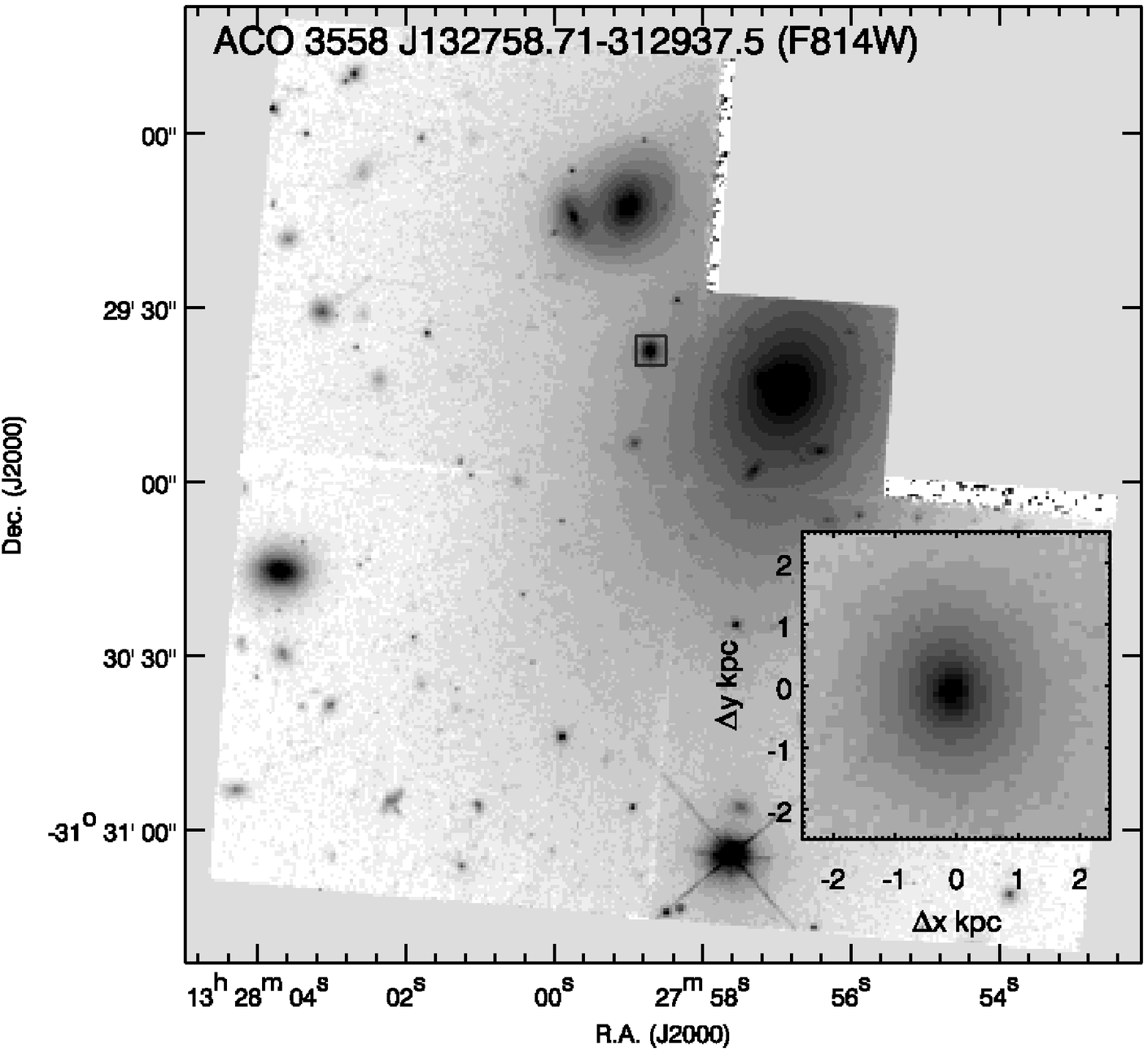}
\includegraphics[width=0.32\hsize]{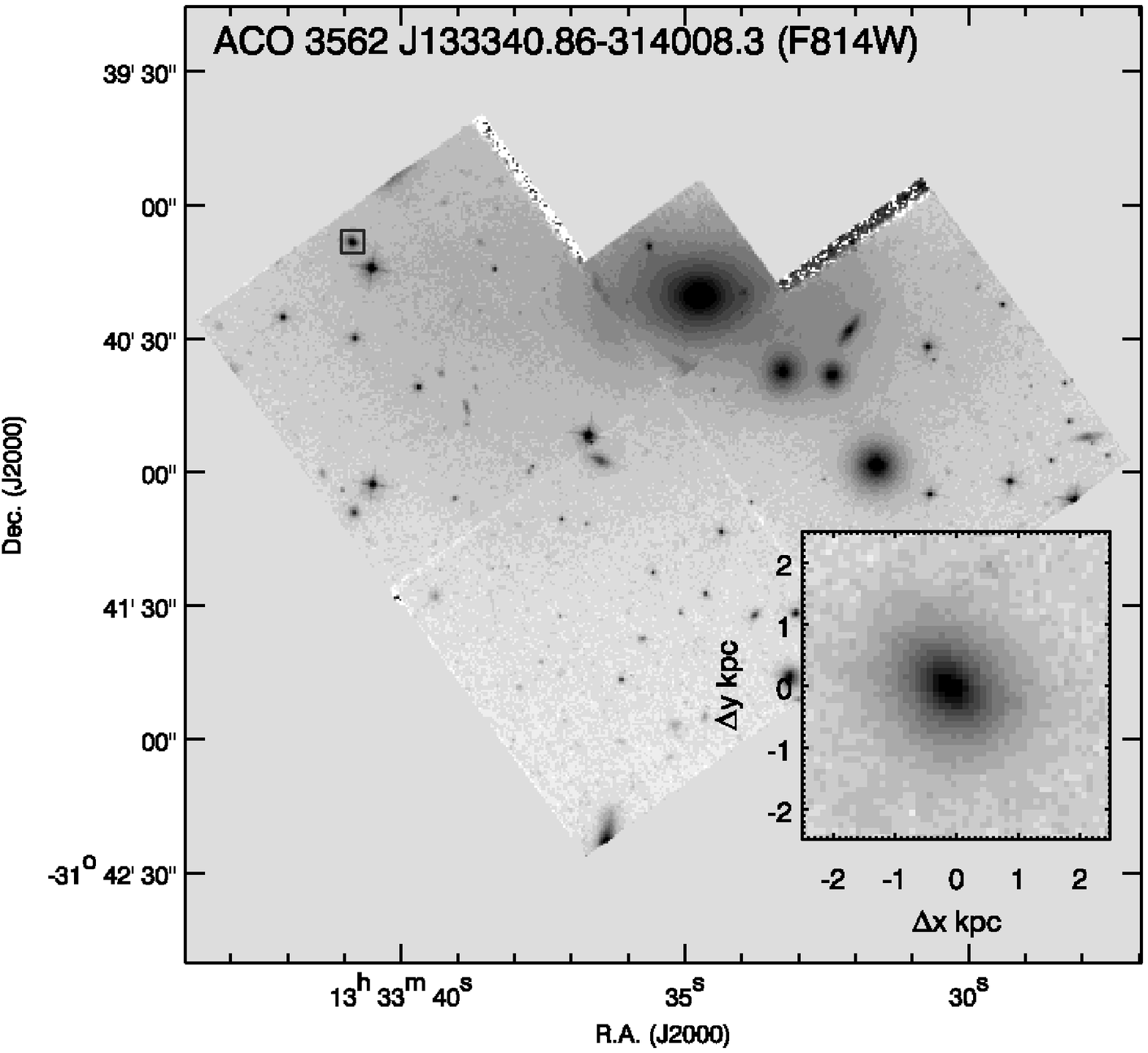}
\includegraphics[width=0.32\hsize]{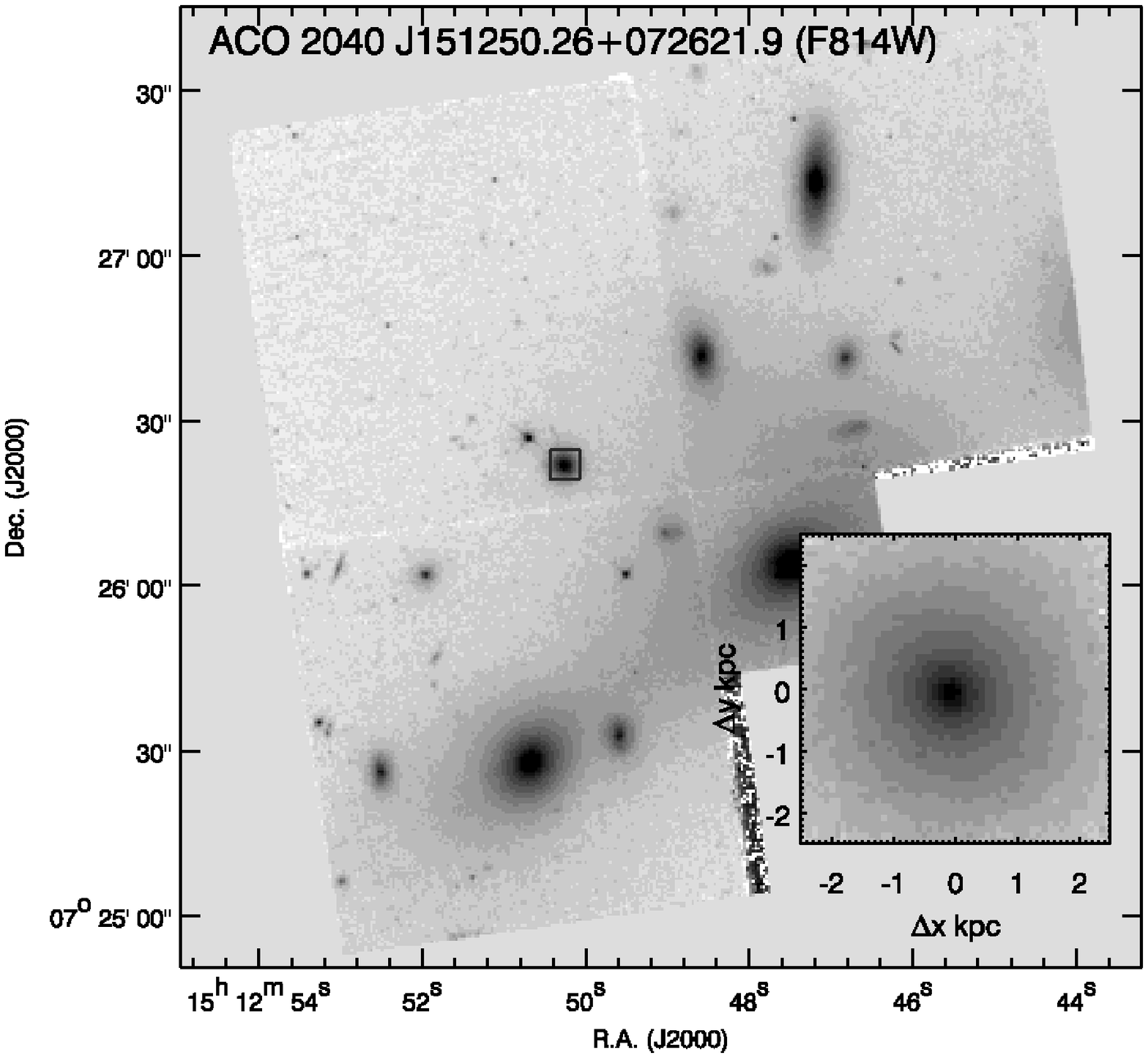}\\
\includegraphics[width=0.32\hsize]{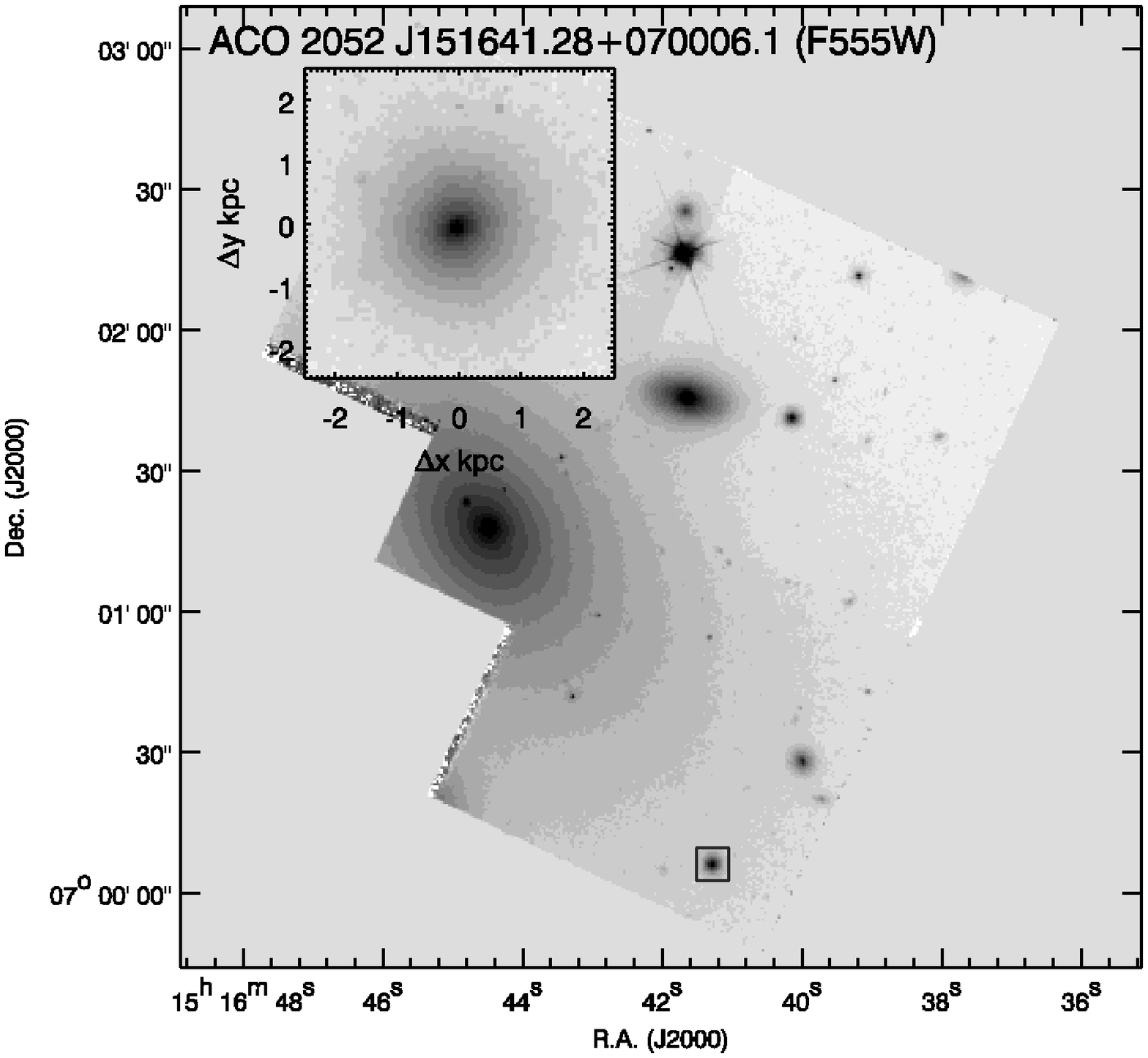}
\includegraphics[width=0.32\hsize]{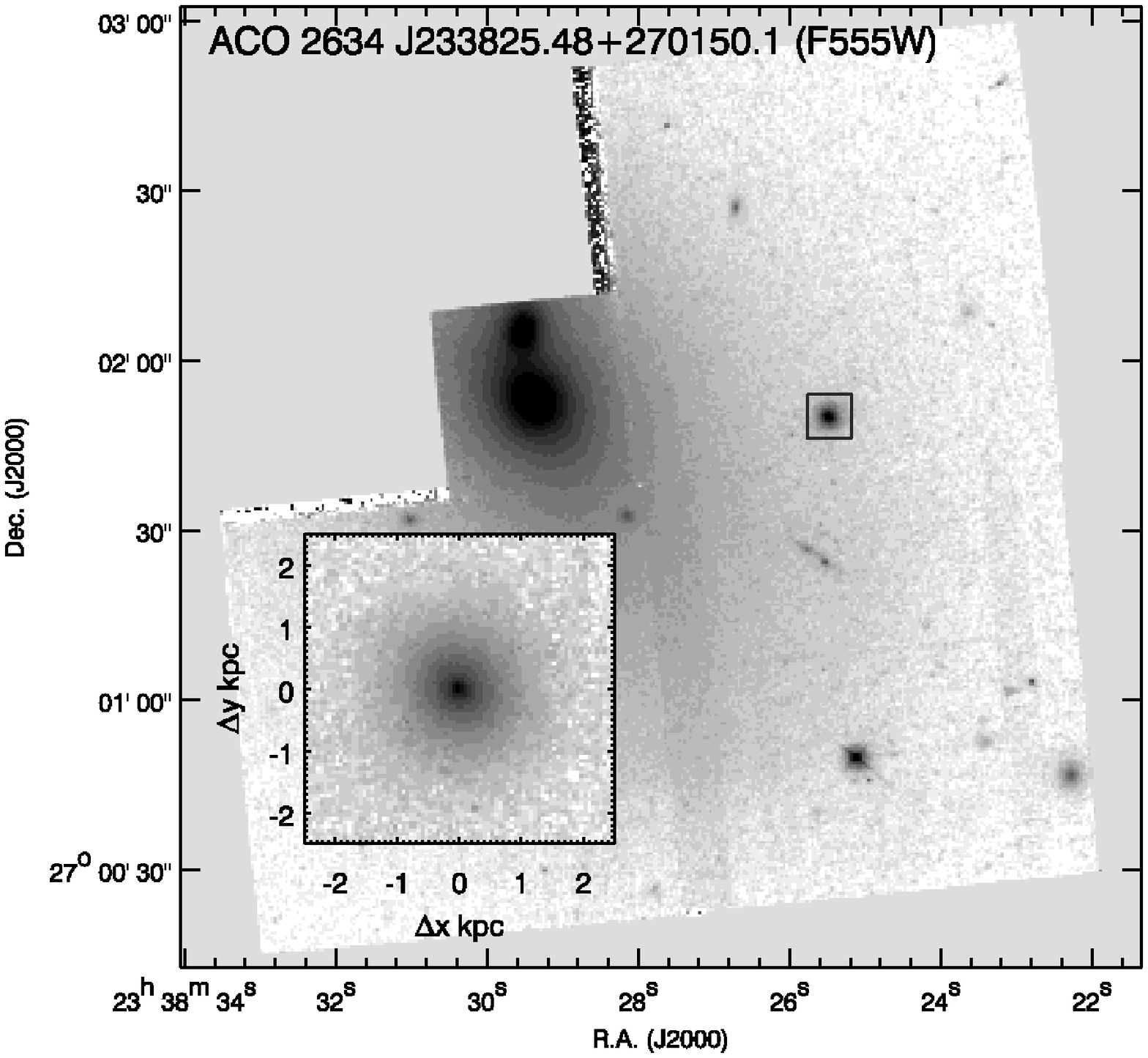}
\includegraphics[width=0.32\hsize]{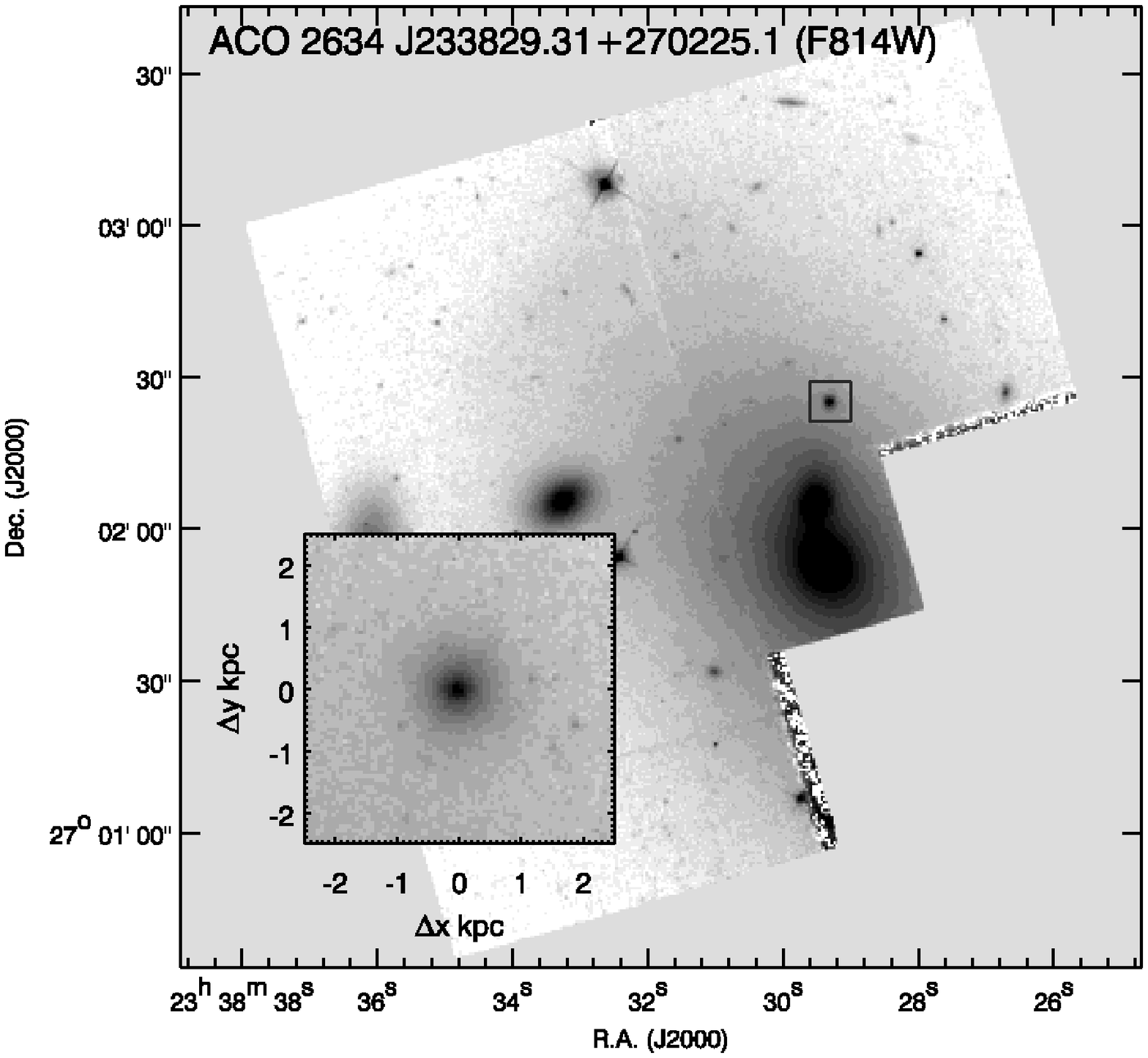}\\
\caption{HST WFPC2 images of cE galaxies -- contd.}
\end{figure}

\begin{figure}
\includegraphics[width=\hsize]{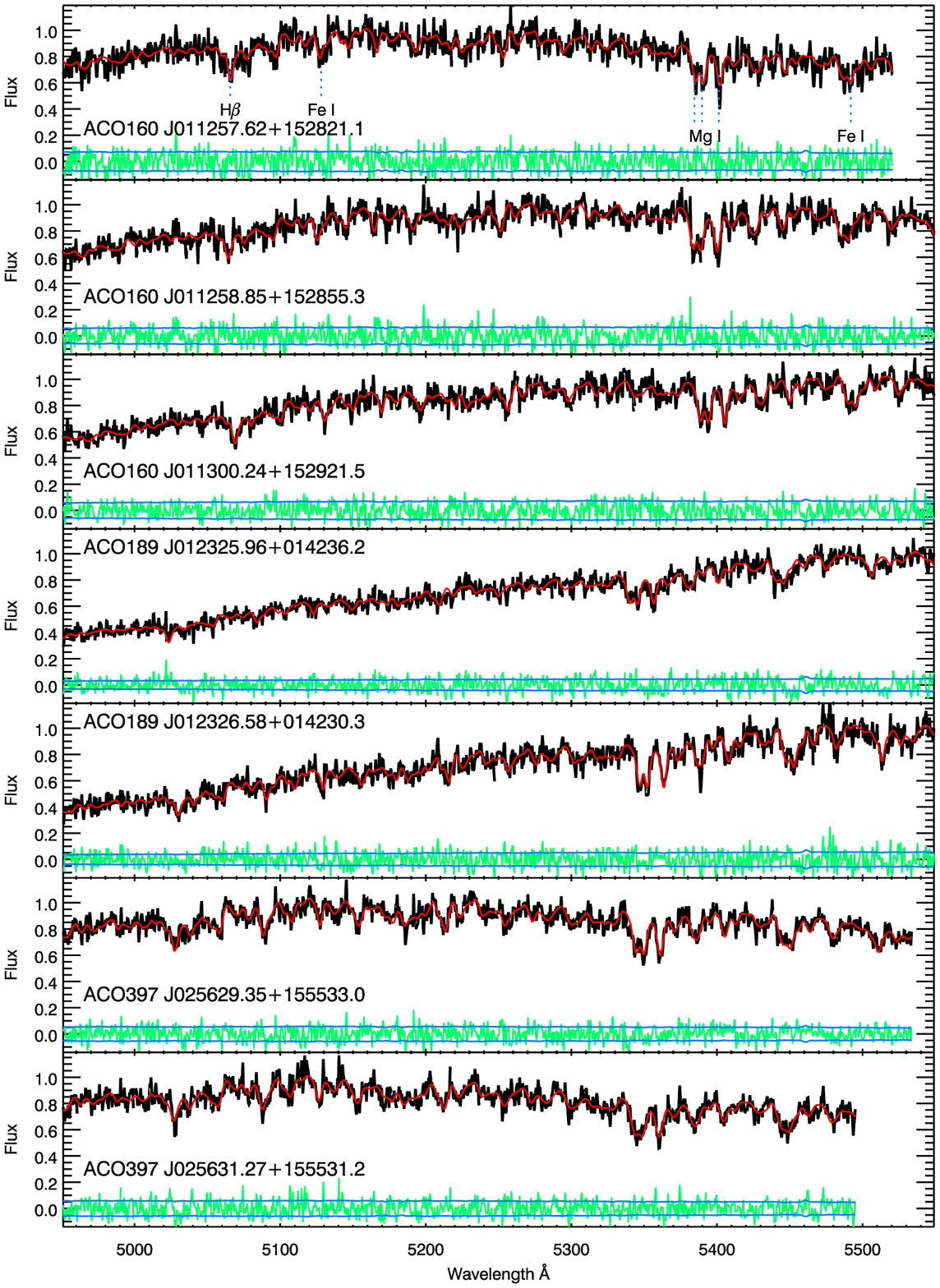}
\caption{BTA spectra of 7 cE galaxies in Abell~160, Abell~189, and Abell~397.
The best-fitting templates and fitting residuals are shown as red and green
solid lines respectively. Flux uncertainties are shown in light blue\label{specBTA}}
\end{figure}

\end{document}